\documentstyle[12pt,dina4,amsfont12]{article} 
 
\input psfig.sty 
\begin{document} 
\newcommand{\pl}[2]{\frac{\partial#1}{\partial#2}} 
\newcommand{\Ta}{{\cal A}} 
\newcommand{\Tb}{{\cal B}} 
\newcommand{\Te}{{\cal E}} 
\newcommand{\bu}{{\bf u} } 
\newcommand{\p}{\partial} 
\newcommand{\og}{\omega} 
\newcommand{\Og}{\Omega} 
\newcommand{\fl}[2]{\frac{#1}{#2}} 
\newcommand{\dt}{\delta} 
\newcommand{\tm}{\times} 
\newcommand{\sm}{\setminus} 
\newcommand{\nn}{\nonumber} 
\newcommand{\ap}{\alpha} 
\newcommand{\bt}{\beta} 
\newcommand{\ld}{\lambda} 
\newcommand{\Gm}{\Gamma} 
\newcommand{\gm}{\gamma} 
\newcommand{\vp}{\varphi} 
\newcommand{\tht}{\theta} 
\newcommand{\ift}{\infty} 
\newcommand{\vep}{\varepsilon} 
\newcommand{\ep}{\epsilon} 
\newcommand{\kp}{\kappa} 
\newcommand{\Dt}{\Delta} 
\newcommand{\Sg}{\Sigma} 
\newcommand{\fa}{\forall} 
\newcommand{\sg}{\sigma} 
\newcommand{\ept}{\emptyset} 
\newcommand{\btd}{\nabla} 
\newcommand{\btu}{\Delta} 
\newcommand{\tg}{\triangle} 
\newcommand{\Th}{{\cal T}_h} 
\newcommand{\ged}{\qquad \Box} 
\newcommand{\bgv}{\Bigg\vert} 
\renewcommand{\theequation}{\arabic{section}.\arabic{equation}} 
\newcommand{\be}{\begin{equation}} 
\newcommand{\ee}{\end{equation}} 
\newcommand{\ba}{\begin{array}} 
\newcommand{\ea}{\end{array}} 
\newcommand{\bea}{\begin{eqnarray}} 
\newcommand{\eea}{\end{eqnarray}} 
\newcommand{\beas}{\begin{eqnarray*}} 
\newcommand{\eeas}{\end{eqnarray*}} 
\newcommand{\dpm}{\displaystyle} 
\newtheorem{theorem}{Theorem}[section] 
\newtheorem{lemma}{Lemma}[section] 
\newtheorem{remark}{Remark}[section] 
\newcommand{\Gmu}{\Gm_{_U}} 
\newcommand{\Gml}{\Gm_{_L}} 
\newcommand{\Gme}{\Gm_e} 
\newcommand{\Gmi}{\Gm_i} 
\newcommand{\lN}{{_N}} 
\newcommand{\tld}[1]{\~{#1}} 
\newcommand{\td}[1]{\tilde{#1}} 
\newcommand{\um}{\mu} 
\newcommand{\bx}{{\bf x} } 
 
\title{Numerical Solution of the Gross-Pitaevskii Equation for 
Bose-Einstein Condensation} 
\author{ {\it Weizhu Bao} 
\thanks{Email address:  bao@cz3.nus.edu.sg. Fax: 65-67746756}\\ 
Department of Computational Science\\ 
National University of Singapore, Singapore 117543.\\ 
\\ 
{\it Dieter Jaksch } 
\thanks{Email address: d.jaksch1@physics.ox.ac.uk.}\\ 
Institut f\"{u}r Theoretische Physik, Universit\"at Innsbruck,\\ 
A--6020 Innsbruck, Austria.\\ 
\\ 
{\it Peter A. Markowich} 
\thanks{Email address: peter.markowich@univie.ac.at, 
http://mailbox.univie.ac.at/peter.markowich.}\\ 
Institute of  Mathematics, University of Vienna\\ 
Boltzmanngasse 9, A-1090 Vienna, Austria.\\ 
} 
 
\date{} 
\maketitle 
 
\begin{abstract} 
We study the numerical solution of the time-dependent 
Gross-Pitaevskii equation (GPE) describing a Bose-Einstein 
condensate (BEC) at zero or very low temperature. In preparation 
for the numerics we scale the 3d Gross-Pitaevskii equation and 
obtain a four-parameter model. Identifying `extreme parameter regimes', 
the model is accessible to analytical perturbation theory, which 
justifies formal procedures well known in the physical literature: 
reduction to 2d and 1d GPEs, 
approximation of ground state solutions of the GPE and geometrical 
optics approximations. Then we use a time-splitting spectral 
method to discretize the time-dependent GPE. Again, perturbation 
theory is used to understand the discretization scheme and to 
choose the spatial/temporal grid in 
dependence of the perturbation parameter. Extensive numerical 
examples in 1d, 2d and 3d for weak/strong interactions, 
defocusing/focusing nonlinearity, and zero/nonzero initial phase 
data are presented to demonstrate the power of the numerical 
method and to discuss the physics of Bose-Einstein condensation. 
 
\end{abstract} 
 

\section{Introduction}\label{si} 
\setcounter{equation}{0} 
 
Recent experimental advances in achieving and observing 
Bose-Einstein condensation (BEC) in trapped neutral atomic vapors 
\cite{Anderson,Ketterle,Bradley} have spurred great excitement in the 
atomic physics community and renewed the interest in studying the 
collective dynamics of macroscopic ensembles of atoms occupying the same 
one-particle quantum state \cite{General,Stringari,Griffin}. The condensate 
typically consists of a few thousands 
 to millions of atoms which are confined 
by a trap potential. In fact, beside the effects of the internal interactions 
between the atoms, the macroscopic behavior of BEC matter is highly 
sensitive to the shape of this external trapping potential. Theoretical 
predictions of the properties of a BEC like the density profile 
\cite{Baym}, collective excitations \cite{colltheo} and the formation 
of vortices \cite{vortextheo} can now be compared with experimental data 
\cite{Anderson,collexp,vortexexp}. Needless to say that this dramatic 
progress on 
the experimental front has stimulated a wave of activity on both the 
theoretical and the numerical front. 
 
The properties of a BEC at temperatures $T$ much smaller than the 
critical condensation 
temperature $T_c$ \cite{LL} are usually well modeled by a 
nonlinear Schr\"{o}dinger equation (NLSE) for the macroscopic wave function 
\cite{LL,Griffin} known as the Gross-Pitaevskii 
equation (GPE) \cite{Gross,Pit}, 
which incorporates the trap potential as well as the interactions among the 
atoms. The effect of the interactions is described by a mean field which 
leads to a nonlinear term in the GPE. The cases of repulsive and attractive 
interactions - which can both be realized in the experiment - correspond 
to defocusing and focusing nonlinearities in the GPE, respectively. Note 
that equations very similar to the GPE also appear in nonlinear optics where 
an index of refraction, which depends on the light intensity, leads to a 
nonlinear term like the one encountered in the GPE. 
 
There has been a series of recent studies which deals with the numerical 
solution of the time-independent GPE for the ground state and the 
time-dependent 
GPE for finding the dynamics of a BEC. Bao et al.~\cite{Bao} presented 
a general 
method to compute the ground state solution via 
directly minimizing  the energy functional and used it 
to compute the ground state of the GPE in different cases. 
Edwards et al.
introduced a Runge-Kutta type method and employed it 
to solve the spherically symmetric 
time-independent GPE \cite{Edwards}. Adhikari~\cite{Adh,Adh1} 
used this approach to obtain the 
ground state solution of the GPE in 2d with radial symmetry. 
Other approaches 
include a finite difference method proposed by Chiofalo 
 et al.~\cite{Tosi} and a 
simple analytical method proposed by Dodd \cite{Dodd}. 
For the numerical solution 
of the time-dependent GPE only  few methods are available, a 
particle-inspired scheme 
proposed by Cerimele et al.~in \cite{Tosi2} and 
a finite difference method used by 
Ruprecht et al.~\cite{Rup}. 
 
In this paper, we take the 3d Gross-Pitaevskii equation, make it 
dimensionless to obtain a four-parameter model, use (singular) 
perturbation theory to discuss semiclassical asymptotics, to 
approximately reduce it to a 2d GPE and a 1d GPE in certain 
limits, and discuss the approximate ground state solution of the 
GPE in two extreme regimes: (very) weak interactions and strong 
repulsive interactions, again using perturbation methods. 
Numerical computations for similar physical
set ups can to a 
large extent be found in the physical 
literatures cf. \cite{Baym}, however they are included here since 
perturbation theory gives a systematic way to obtain rigorously 
(validate) these approximations and since they are used as important 
preparatory steps for the numerical simulations (approximate 
ground states usually serve as initial data). Then we use the 
time-splitting spectral method, which was studied in Bao et 
al.~\cite{Bao1,Bao2} for the Schr\"{o}dinger equation in the 
semiclassical regime, to discretize the time-dependent GPE. The 
merit of the numerical method is that it is explicit, 
unconditionally stable, time reversible, time-transverse 
invariant, and conserves the  position density. In fact, the 
spectral method has shown great success in solving problems 
arising from many areas of science \cite{spectral,Canuto} due to 
its spatially spectral accuracy. The split-step procedure was 
presented for differential equations in \cite{Strang} and applied 
to Schr\"{o}dinger equations \cite{Hardin,Taha1,Fornberg} and the 
KDV equation \cite{Taha2}, as well as used with an iterative 
procedure for optical fibers \cite{Agrawal}. In this paper we 
perform grid control using singular 
perturbation theory and present extensive numerical examples in 
1d, 2d and 3d for weak/strong interactions, defocusing/focusing 
nonlinearity, and zero/nonzero initial phase data. 
 
The paper is organized as follows. In section \ref{sgpel} we start 
out with the 3d GPE, scale it to get a four-parameter model, show 
how to reduce it to lower dimensions and give the approximate 
ground state solution in the two mentioned extreme regimes of 
(very) weak interactions and strong repulsive interactions and 
discuss semiclassical asymptotics. In section \ref{sna} we present 
the time-splitting spectral method for the GPE. In section 
\ref{sne} numerical tests of the GPE for different cases including 
weak/strong interactions, defocusing/focusing nonlinearity, and 
zero/nonzero initial phase data are presented. In section \ref{sc} 
a summary is given. 
 
\section{Gross-Pitaevskii equation}\label{sgpel} 
\setcounter{equation}{0} 
 
At temperatures $T$ much smaller than the critical temperature 
$T_c$ \cite{LL}, a BEC is well described by the macroscopic wave function 
 $\psi = \psi({\bf x},t)$ whose evolution is governed by 
a self-consistent, 
mean field nonlinear Schr\"{o}dinger equation (NLSE) 
known as the  Gross-Pitaevskii equation \cite{Gross,Pit}. 
If a harmonic trap potential is considered, the equation becomes 
\be 
\label{gpe1} 
i\hbar\pl{\psi(\bx,t)}{t}=-\fl{\hbar^2}{2m}\btd^2 \psi(\bx,t)+ 
\fl{m}{2}\left(\og_x^2 x^2+\og_y^2 y^2+\og_z^2 z^2\right)\psi(\bx,t) 
+N U_0 |\psi(\bx,t)|^2\psi(\bx,t), 
\ee 
where $\bx=(x,y,z)^T$ is the spatial coordinate vector, 
$m$ is the atomic mass, $\hbar$ is the Planck constant, 
$N$ is the number of atoms in the condensate, 
and $\og_x$, $\og_y$ and $\og_z$ are the trap 
frequencies in $x$, $y$ and $z$-direction, respectively. 
For the following we assume (w.r.o.g.) $\og_x\le\og_y\le\og_z$. 
When $\og_x=\og_y=\og_z$, the trap potential is isotropic. 
$U_0$ describes the interaction between atoms in the 
condensate and has the form 
\be 
\label{U0} 
U_0=\fl{4\pi \hbar^2 a}{m}, 
\ee 
where $a$ is the $s$-wave scattering length (positive for repulsive 
interaction and negative for attractive interaction). It is necessary 
to ensure that the wave function is properly normalized. 
Specifically, we require 
\be 
\label{norm} 
\int_{{\Bbb R}^3} \; |\psi(\bx,t)|^2\;d\bx=1. 
\ee 
A typical set of parameters used in current experiments 
with ${}^{87}$Rb is given 
by 
\be 
\label{numb} 
m=1.44\tm 10^{-25}[kg], \ 
\og_x=\og_y=\og_z=20\pi[rad/s], \ 
a=5.1\tm 10^{-9}[m], \ N:\ 10^2\sim 10^7 
\ee 
and the Planck constant has the value 
\[ \hbar =1.05\tm 10^{-34}\ [Js].\]

\subsection{Dimensionless GPE} 
 
In order to scale the equation (\ref{gpe1}) under the 
normalization (\ref{norm}),  we introduce \be \label{dml} 
\tilde{t}=\og_x t, \qquad \tilde{\bx}=\fl{\bx}{x_s}, \qquad 
\tilde{\psi}(\tilde \bx,\tilde t)=x_s^{3/2} \psi(\bx,t), \ee where 
$x_s$ is the characteristic `length' of the condensate. Plugging 
(\ref{dml}) into (\ref{gpe1}), multiplying  by $\fl{1}{m \og_x^2 
x_s^{1/2}}$, 
 and then removing all \~{}, we 
obtain the following dimensionless Gross-Pitaevskii equation 
under the normalization (\ref{norm}) in three spatial dimensions 
\be 
\label{gpe2} 
i\vep \pl{\psi(\bx,t)}{t}=-\fl{\vep^2}{2}\btd^2 \psi(\bx,t)+ 
V(\bx)\psi(\bx,t) 
+ \dt \vep^{5/2} |\psi(\bx,t)|^2\psi(\bx,t), 
\ee 
where 
\beas 
\label{poten} 
&&V(\bx)=\fl{1}{2}\left(x^2+\gm_y^2 y^2+\gm_z^2 z^2\right),\\ 
&&\vep=\fl{\hbar}{\og_x m x_s^2}=\left(\fl{a_0}{x_s}\right)^2,\qquad \qquad 
\gm_y=\fl{\og_y}{\og_x}, \qquad \qquad \gm_z=\fl{\og_z}{\og_x},\\ 
&&\dt=\fl{U_0 N}{a_0^3\hbar \og_x}=\fl{4\pi aN}{a_0}, \qquad 
a_0=\sqrt{\fl{\hbar}{\og_x m}}, 
\eeas 
with $a_0$ the length of the harmonic oscillator ground state 
(in $x$-direction). The 
coefficient of the nonlinearity 
 of (\ref{gpe2}) (interaction strength parameter) 
can also be expressed as 
\be 
\label{coffn} 
\kp:=\dt \vep^{5/2}=\fl{4\pi a N}{a_0}\left(\fl{a_0}{x_s}\right)^5 
=\fl{1}{2}\fl{8\pi aN}{x_s^3}\fl{a_0^4}{x_s^2} 
=\fl{{\rm sgn}(a)}{2}\fl{a_0^2}{x_h^2}\fl{a_0^2}{x_s^2} 
=\fl{{\rm sgn}(a)}{2}\left(\fl{a_0}{x_h}\fl{a_0}{x_s}\right)^2, 
\ee 
where $x_h$ is the healing length \cite{Baym} with 
\be 
\label{healing} 
x_h:=\left(\fl{8\pi |a| N}{x_s^3}\right)^{-\fl{1}{2}}. 
\ee 
If we plug the typical set of parameter values (\ref{numb}) 
into the 
above parameters, we find 
\[a_0\approx 0.3407\tm 10^{-5} [m], \qquad \dt\approx 
0.01881 N: \ 1.881 \sim 188100.\] 
 
\begin{remark}\label{vep1} 
If one chooses $x_s=a_0$ in (\ref{dml}), then $\vep=1$ in (\ref{gpe2}), 
 $\kp=\dt$ in (\ref{coffn}), and the  equation 
(\ref{gpe2}) takes the form often appearing in the physical 
literature. This choice for $x_s$ is appropriate in the weak 
interaction regime characterized by $4\pi|a|N\ll a_0$ 
and in the moderate interaction regime where 
$4\pi|a|N\approx a_0$. In the strong interaction regime 
$4\pi|a|N\gg a_0$ a different choice is 
 more appropriate, namely $x_s=(4\pi|a|Na_0^4)^{1/5}$, 
which gives $|\kp|=1$ and $\vep=\left(\fl{a_0}{4\pi|a|N}\right)^{1/5}\ll1$. 
Other choices for $x_s$, based on 
approximating the actual condensate length scale,  shall be 
discussed in Section 2.2. Note that the choice of $x_s$ determines 
the observation scale of the condensate and decides: (i) which 
phenomena are `visible' by asymptotic analysis, (ii) which 
phenomena can be resolved by discretization on specified 
spatial/temporal grids. 
\end{remark} 
 
\bigskip 
 
Thus, there are two extreme regimes: one is when $\vep=O(1)$ 
($\Leftrightarrow a_0=O(x_s)$) and $\kp=\dt \vep^{5/2}=o(1)$ 
($\Leftrightarrow 4\pi |a|N\ll a_0$), then equation (\ref{gpe2}) 
describes a weakly interacting condensate. The other is when 
$\vep=o(1)$ ($\Leftrightarrow x_s\gg a_0$) and 
$\kp=\dt\vep^{5/2}=O(1)$ (implying $4\pi |a|N\gg a_0$) (or 
$\vep=1$ and $\kp=\dt \vep^{5/2}=\dt $ with $|\dt|\gg1$ by the 
rescaling $\bx\to\vep^{1/2}\bx$, $\psi\to \psi/\vep^{3/4}$), then 
(\ref{gpe2}) corresponds to a strongly interacting condensate 
(Thomas-Fermi regime \cite{Baym}) or, equivalently, to the 
semiclassical regime. We recall that the equation (\ref{gpe2}) is 
\underline{regularly perturbed} in the case of weak interactions 
and \underline{singularly perturbed} in the semiclassical regime. 
Analytical techniques of asymptotic analysis are available in both 
cases providing structural information on the solutions of 
(\ref{gpe2}) and on numerical discretization schemes 
(spatial/temporal grid control, control of the computational 
domain, error estimates in linearized cases). 
 
\subsection{Approximate ground state solution in 3d} 
 
To find a stationary state of (\ref{gpe2}), we write 
\be 
\label{stat} \psi(\bx,t)=e^{-i \mu t / \vep} \phi(\bx), 
\ee 
where $\mu$ is the chemical potential of the condensate. Inserting 
into (\ref{gpe2}) gives the following equation for $\phi(\bx)$ 
\be 
\label{gpeg1} 
\mu \phi(\bx)=-\fl{\vep^2}{2}\btd^2 \phi(\bx)+ 
V(\bx)\phi(\bx) + \kp |\phi(\bx)|^2\phi(\bx), \qquad \bx\in {\Bbb 
R}^3, 
\ee 
under the normalization condition 
\be \label{normgg} 
\int_{{\Bbb R}^3} \; |\phi(\bx)|^2\;d\bx=1. 
\ee 
This is a 
nonlinear eigenvalue problem. The Bose-Einstein condensate 
ground-state wave function $\phi_g(\bx)$ is found by solving this 
eigenvalue problem under the normalization condition 
(\ref{normgg}) with the minimal chemical potential $\mu_g$. 
Usually, the ground state problem is formulated variationally. 
 Define the energy functional 
\begin{equation} 
E(\phi):=\frac{\vep^2}{2} \int_{{\Bbb R}^3} 
\left|\nabla \phi\right|^2 d\bx 
+ \int_{{\Bbb R}^3} V(x) \left|\phi\right|^2 d\bx + \frac{\kappa}{2} 
\int_{{\Bbb R}^3} \left|\phi\right|^4 d\bx. 
\end{equation} 
It is easy to see that critical points of $E$ are `eigenfunctions' of the 
nonlinear Hamiltonian. To compute the ground state $\phi_g$ we solve 
\begin{equation} 
E(\phi_g)=\min_{\int_{{\Bbb R}^3} \; |\phi|^2\;d\bx=1}\ E(\phi), 
\qquad \um_g=E(\phi_g)+\fl{\kp}{2}\int_{{\Bbb R}^3} 
\left|\phi_g\right|^4 d\bx. 
\end{equation} 
In the case of a defocusing (stable) condensate the energy 
functional $E(\psi)$ is positive, coercive and weakly lower 
semicontinuous on the unit sphere in $L^2({\Bbb R}^3)$, thus the 
existence of a minimum follows from standard theory. For understanding 
the uniqueness question note that $E(\ap \phi_g)=E(\phi_g)$ for 
all $\ap\in {\Bbb C}$ with $|\ap|=1$. Thus an additional 
constraint has to be introduced to show uniqueness, e.g. $\phi_g$ real 
valued and $\phi_g(\bx)>0$ for all $\bx\in{\Bbb R}^3$ (see \cite{Lieb}). 
For focusing (unstable) 3-dimensional condensates the energy functional 
$E(\psi)$ is not bounded from below on the unit sphere of 
$L^2({\Bbb R}^3)$. Thus, an absolute minimum of $E(\psi)$ does 
not exist on $\{\psi\in L^2({\Bbb R}^3) \ |\ \|\psi\|_{L^2}^2=1\}$. 
The interpretation of critical points (local minimum, saddle points 
obtained by min-max-theory) as physically relevant 
ground states is by no means clear. 
 
 Using  (simple) perturbation methods, we present here the approximate 
ground state solution of (\ref{gpe2}) in the two extreme regimes 
of weak repulsive or attractive interactions and strong repulsive 
interactions (see \cite{Baym} for a discussion in physical 
literature). 
 
These approximate ground state solutions are used in reducing the 
3d GPE to a 2d GPE and a 1d GPE - see the next subsection for 
details - and as initial data for the numerical solution of the 
time-dependent GPE in section \ref{sne} (see the subsequent discussion). 
 
For a weakly interacting condensate, i.e. $\vep=O(1)$ and 
$\kp=o(1)$, we drop the nonlinear term (i.e. the last term on the 
right hand side of (\ref{gpe2})) and find the harmonic oscillator 
equation 
 \be \label{gpegw} 
\mu \phi(\bx)=-\fl{\vep^2}{2}\btd^2 \phi(\bx)+ 
\fl{1}{2}\left(x^2+\gm_y^2 y^2+\gm_z^2 z^2\right)\phi(\bx). 
 \ee 
The ground state solution of (\ref{gpegw}) is 
 \be \label{gssw} 
\mu_g^w=\fl{1+\gm_y+\gm_z}{2} \vep, \qquad 
\phi_g^w(\bx)=\fl{(\gm_y \gm_z)^{1/4}}{(\pi\vep)^{3/4}} \; 
e^{-(x^2+\gm_y y^2+\gm_z z^2)/2\vep}. 
 \ee 
It can be viewed as an approximate ground state solution of 
(\ref{gpe2}) in the case of a weakly interacting condensate, with 
an $O(\kp)$-error in approximating the chemical potential and 
ground state. From (\ref{gssw}), we can see that the diameter of 
the ground state solution (computed according to the 
formula (\ref{cw})) in the weakly interacting condensate is 
\[ x_g^w=O\left(\sqrt{\vep}\right) 
=O(1)  \qquad (\hbox{after the scaling (\ref{dml})}).\] 
Also we remark that the condensate widths in $y$ and $z$-directions 
of the approximate ground state $\phi_g^w$ are 
$O\left(\sqrt{\vep}/\sqrt{\gm_y}\right)=O\left(1/\sqrt{\gm_y}\right)$ 
and $O\left(\sqrt{\vep}/\sqrt{\gm_z}\right)=O\left(1/\sqrt{\gm_z}\right)$, 
respectively. Clearly, this is important for the control of the 
computational domain. 
 
For a condensate with strong repulsive interactions, i.e. $\vep=o(1)$, 
$\kp=O(1)$ and $\kappa>0$, we drop the diffusion term (i.e. the first term 
on the right hand side of (\ref{gpe2})) corresponding to the Thomas-Fermi 
approximation \cite{Baym}: 
 \be \label{gpegs} 
\mu \phi(\bx)=V(\bx)\phi(\bx) + \kp |\phi(\bx)|^2\phi(\bx), \qquad 
\bx\in {\Bbb R}^3. 
 \ee 
The ground state solution of (\ref{gpegs}) is the compactly supported 
function $\phi_g^s$: 
 \be \label{gss} 
 \mu_g^s=\fl{\vep}{2}\left(\fl{15\dt 
\gm_y\gm_z}{4\pi}\right)^{2/5}= \fl{1}{2}\left(\fl{15\kp 
\gm_y\gm_z}{4\pi}\right)^{2/5}, \ \phi_g^s(\bx)=\left\{\ba{ll} 
\sqrt{\left(\mu_g^s -V(\bx)\right)/\kp}, 
&\ V(\bx)< \mu_g^s,\\ 
0, & \hbox{otherwise}. 
\ea\right. 
 \ee 
This shows that the diameter of the ground state 
solution in the strongly interacting repulsive condensate is 
\[x_g^s=\sqrt{\mu_g^s} =O((\dt\gm_y\gm_z)^{1/5})\fl{a_0}{x_s}= 
O\left(\left(\fl{4\pi |a| N \gm_y\gm_z}{a_0}\right)^{1/5}\right) 
\fl{a_0}{x_s} \ (\hbox{again after the scaling (\ref{dml})}). 
\] 
Approximate widths of $\phi_g^s$ in the $y$ and $z$-directions 
are $O\left((\kp \gm_z)^{1/5}/\gm_y^{4/5}\right)= 
O\left(\gm_z^{1/5}/\gm_y^{4/5}\right)$ 
and $O\left((\kp \gm_y)^{1/5}/\gm_z^{4/5}\right)= 
O\left(\gm_y^{1/5}/\gm_z^{4/5}\right)$, respectively. 
 
This analysis suggests to choose the characteristic 
condensate length $x_s$ such 
that the (dimensionless) ground state width $x_g^w$ is $O(1)$ 
after the scaling (\ref{dml}),  i.e.  $x_s=O(a_0)$ for weak 
interaction (as in Remark 2.1) 
and $x_s=O\left(\left( \fl{4\pi |a| N}{a_0} 
\gm_y\gm_z\right)^{1/5}a_0\right)$ for strong repulsive 
interaction (different from Remark 2.1 if $\gm_y\gg1$ 
or $\gm_z\gg1$). If we use the typical set of parameter values 
(\ref{numb}) 
 in the above identity, we obtain 
\[\vep:\ 0.0078\sim 1.\]

\begin{remark}\label{vepnew} 
The approximate ground state $\phi_g^w$ in the weak-interaction 
regime has finite energy, more precisely 
\begin{equation} 
E(\phi_g^w) = \mu_g^w + \frac{\kappa}{4} \left(\frac{\gamma_y 
\gamma_z}{\pi^3 \vep^3}\right)^{1/2} = \mu_g^w + O(\kappa) 
\end{equation} 
as $\kappa \rightarrow 0$ for $\vep=O(1)$, $\gamma_z =O(1)$, 
$\gamma_y=O(1)$. 
 
Contrarily the energy of the Thomas-Fermi approximation is 
infinite 
\begin{equation} 
E(\phi_g^s) = + \infty 
\end{equation} 
due to the low regularity of $\phi_g^s$ at the free boundary 
$V(x)=\mu_g^s$. More precisely, $\phi_g^s$ is locally 
$C^{1/2}$ at the interface but not $H^1_{\rm loc}({\Bbb R}^3)$. This is a 
typical behavior for solutions of free boundary value problems, 
which indicates that $\phi_g^s$ does not approximate $\phi_g$ to 
the full $O(\vep^2)$ - order, as indicated by formal 
consistency. An interface layer correction has to be constructed 
in order to improve the approximation quality. For a convergence proof of 
$\phi_g^s\to\phi_g$ (without convergence rate) we refer to \cite{Lieb}. 
 
It is of course tempting to use approximate ground states as 
initial data for the GPE when simulating Bose-Einstein 
condensation. In the weak interaction case this produces 
$O(\kappa)$ - errors in time dependent simulations on $O(1)$ time 
intervals. In the strong interaction case an initial wave function 
$\phi_g^s$ produces time - dependent solutions with infinite 
energy (which usually generates breathing modes, cf.~Example 3 III 
in Sec.~\ref{secapp}) and the error in the wave function 
introduced by this is typically significantly larger than 
$O(\vep^2)$. 
\end{remark} 
 
\bigskip 
 
\subsection{Reduction to lower dimensions} 
 
In two important cases, the 3d Gross-Pitaevskii equation (\ref{gpe2}) 
can approximately be reduced to a lower dimensional PDE. 
For a disk-shaped condensate with small height, i.e. 
 \be 
\label{r2d} 
\og_x\approx \og_y, \quad \og_z\gg \og_x, \qquad \Longleftrightarrow \qquad 
\gm_y\approx1, \quad \gm_z\gg 1, 
 \ee 
the 3d GPE (\ref{gpe2}) can be reduced to a 2d  GPE with $\bx=(x,y)^T$ by 
assuming that the time evolution does not cause excitations along the 
$z$-axis since these have a large energy of approximately 
$\hbar \og_z$ compared to 
excitations along the $x$ and $y$-axis with energies of about $\hbar \og_x$. 
To understand this, consider the total condensate energy $E[\psi(t)]$: 
 \bea \label{energy} 
E[\psi(t)]&=&\fl{\vep^2}{2}\int_{{\Bbb R}^3}|\btd\psi(t)|^2d\bx 
+\fl{1}{2}\int_{{\Bbb R}^3} \left(x^2+\gm_y^2 
y^2\right)|\psi(t)|^2d\bx \nn\\ 
&&+\fl{\gm_z^2}{2}\int_{{\Bbb R}^3} 
z^2|\psi(t)|^2d\bx+\frac{\kappa}{2} \int_{{\Bbb R}^3} |\psi(t)|^4 
d\bx. 
 \eea 
Multiplying (\ref{gpe2}) by $\overline{\psi_t}$ and integrating by 
parts show the energy conservation 
 \be 
\label{energyc} 
E[\psi(t)]=E[\psi_I], \qquad \forall t, 
 \ee 
where $\psi_I=\psi(t=0)$ is the initial function which may depend 
on all parameters $\vep$, $\gm_y$, $\gm_z$ and $\kp$. Now assume 
that $\psi_I$ satisfies 
 \be \label{energyl} 
\fl{E[\psi_I]}{\gm_z^2}\to 0, \qquad \hbox{as} \quad \gm_z\to 
\ift. 
 \ee 
Take a sequence $\gm_z\to\ift$ (and keep all other parameters 
fixed). Since $\dpm\int_{{\Bbb R}^3}|\psi(t)|^2\;d\bx=1$ we 
conclude from weak compactness that there is a positive measure 
$n^0(t)$ such that 
\[|\psi(t)|^2 \rightharpoonup n^0(t) 
\quad \hbox{weakly as}\quad \gm_z\to\ift.\] 
Energy conservation implies 
\[\int_{{\Bbb R}^3}\; z^2|\psi(t)|^2\;d\bx \to 0, \quad \hbox{as} 
\quad \gm_z\to\ift 
\] 
and thus we conclude concentration of the condensate in the plane 
$z=0$: 
\[n^0(x,y,z,t)=n_2^0(x,y,t)\dt(z),\] 
where $n_2^0(t)$ is a positive measure on ${\Bbb R}^2$. 
 
  Now let $\psi_3=\psi_3(z)$ be a wave-function with 
\[\int_{{\Bbb R}} \; |\psi_3(z)|^2\;dz =1,\] 
depending on $\gm_z$ such that 
 \be 
 |\psi_3(z)|^2 \rightharpoonup 
\dt(z), \qquad \hbox{as} \quad \gm_z\to \ift, 
 \ee 
Denote by $S$ the subspace 
\[S=\{\psi=\psi_2(x,y)\psi_3(z)\; |\; \psi_2\in L^2({\Bbb R}^2)\}\] 
and let 
\[\Pi:\; L^2({\Bbb R}^3)\to S\subseteq L^2({\Bbb R}^3)\] 
be the projection on $S$: 
\[ (\Pi\psi)(x,y,z)=\psi_3(z)\int_{{\Bbb R}} \; \overline{\psi_3}(\sg) 
\; \psi(x,y,\sg)\; d\sg.\] 
Now write the equation (\ref{gpe2}) in the form 
\[i\vep \psi_t={\cal A}\psi +{\cal F}(\psi),\] 
where ${\cal A}\psi$ stands for the linear part and ${\cal 
F}(\psi)$ for the nonlinearity. Applying $\Pi$ to the GP-equation 
gives \bea \label{proj} 
i\vep (\Pi \psi)_t&=&\Pi {\cal A} \psi +\Pi {\cal F}(\psi)\nn\\ 
&=&\Pi {\cal A} (\Pi\psi) +\Pi {\cal F}(\Pi\psi) +\Pi\left((\Pi 
{\cal A}-{\cal A}\Pi)\psi+(\Pi{\cal F}(\psi)- {\cal F}(\Pi 
\psi))\right). \qquad 
 \eea 
 The projection approximation of (\ref{gpe2}) 
is now obtained by dropping the commutator terms. it reads \bea 
\label{proj2j} 
&&i\vep (\Pi \sg)_t=\Pi {\cal A} (\Pi\sg) +\Pi {\cal F}(\Pi\sg),\\ 
\label{proj3j} &&(\Pi\sg)(t=0)=\Pi \psi_I, \eea or explicitly, 
with 
 \be \label{proj4j} 
 (\Pi \sg)(x,y,z,t)=:\psi_2(x,y,t)\psi_3(z), 
 \ee 
we find 
 \be \label{gpe2d} 
 i\vep 
\pl{\psi_2}{t}=-\fl{\vep^2}{2}\btd^2 \psi_2+ 
\fl{1}{2}\left(x^2+\gm_y^2 y^2 + C \right) \psi_2 + \left(\dt 
\vep^{5/2}\int_{-\ift}^\ift \psi_3^4(z)\,dz\right) 
|\psi_2|^2\psi_2, \ee where 
\[C=\gm_z^2\int_{-\ift}^\ift \; z^2|\psi_3(z)|^2\;dz+ 
\vep^2\int_{-\ift}^\ift\; \left|\fl{d\psi_3}{dz}\right|^2\;dz.\] 
Since this GPE is time-transverse invariant, we can replace 
$\psi_2\to \psi\;e^{-i\fl{C}{2\vep}}$ and drop the constant $C$ in 
the trap potential. The observables are not affected by this. 
 
The `effective' GP-equation (\ref{gpe2d}) is well known in the 
physical literature \cite{BeyondGPE}, where the projection method 
is often referred to as `integrating out the $z$-coordinate'. 
However, an analysis of the limit process $\gm_z\to\ift$ has to be 
based on the derivation as presented above, in particular on 
studying the commutators $\Pi {\cal A}-{\cal A}\Pi$, $\Pi {\cal F} 
-{\cal F}\Pi $. In the case of small interaction 
$\vep=O(1)$, $\kp=o(1)$, a good choice for $\psi_3(z)$ is the 
ground state of the harmonic oscillator in $z$-dimension: \be 
\label{psi3} \psi_3(z)=\left(\fl{\gm_z}{\pi \vep}\right)^{1/4}\; 
e^{-\gm_z z^2/(2\vep)}. 
 \ee 
Note that $|\psi_3(z)|^2\rightharpoonup \dt(z)$ as $\gm_z\to \ift$ 
and that $\Pi {\cal A}={\cal A}\Pi$ such that the error in 
approximating $\Pi\psi(t)$ by $\Pi\sg$ is determined by the 
commutator of the nonlinearity, which is $O(\kp)$. 
 
For condensates with other than small interaction the choice of 
$\psi_3$ is much less obvious. Often one assumes  that the 
condensate density along the $z$-axis well described by the 
$(x,y)$-trace of the ground state position density $|\phi_g|^2$ 
  \be 
\label{pps} 
|\psi(x,y,z,t)|^2 \approx |\psi_2(x,y,t)|^2 \int_{{\Bbb R}^2} 
\; |\phi_g(x_1,y_1,z)|^2\;dx_1dy_1 
\ee 
and (taking a pure-state-approximation) 
\be 
\label{pph} 
\psi_3(z)=\left(\int_{{\Bbb R}^2} 
\; |\phi_g(x,y,z)|^2\;dxdy\right)^{1/2} 
\ee 
A mathematical analysis of the limit 
process $\gm_z\to\ift$ is currently under study.

\bigskip 
 
For a cigar-shaped condensate 
\be 
\label{r2dd} 
\og_y\gg \og_x, \quad \og_z\gg \og_x, \qquad \Longleftrightarrow\qquad 
\gm_y\gg1, \quad \gm_z\gg 1, 
 \ee 
the 3d GPE (\ref{gpe2}) can be reduced to a 1d GPE by proceeding 
analogously. 
 
Then the 3d GPE (\ref{gpe2}), 2d GPE and 1d GPE can then be 
written in a unified way 
 \be 
\label{gpeg} 
i\vep \pl{\psi(\bx,t)}{t}=-\fl{\vep^2}{2}\btd^2 \psi(\bx,t)+ 
V_d(\bx)\psi(\bx,t) 
+ \kp_d\; |\psi(\bx,t)|^2\psi(\bx,t), \qquad \bx\in {\Bbb R}^d, 
\ee 
where 
\be 
\label{uf} 
\kp_d=\dt \vep^{5/2}\;\left\{\ba{l} 
\int_{{\Bbb R}^2} \psi_{23}^4(y,z)\;dy dz, \\ 
\int_{{\Bbb R}} \psi_3^4(z)\;dz, \\ 
1,\\ 
\ea\right. 
\quad 
V_d(\bx)=\left\{\ba{ll} 
 \fl{1}{2}x^2,  &\quad d=1, \\ 
 \fl{1}{2}\left(x^2+\gm_y^2 y^2\right), &\quad d=2, \\ 
 \fl{1}{2}\left(x^2+\gm_y^2 y^2+\gm_z^2 z^2\right), &\quad d=3.\\ 
\ea\right. 
\ee 
The normalization condition for (\ref{gpeg}) is 
\be 
\label{normg} 
\int_{{\Bbb R}^d} \; |\psi(\bx,t)|^2\;d\bx=1. 
\ee 
By using the approximate ground state of Section 2.2, 
we derive - after simple calculations - 
for a weakly interacting condensate 
\be 
\label{ufw} 
\kp_d:=\kp_d^w=\left\{\ba{ll} 
 \fl{\kp(\gm_y \gm_z)^{1/2}}{2\pi\vep}= 
\dt \vep^{3/2} \fl{(\gm_y \gm_z)^{1/2}}{2\pi}, &\qquad d=1, \\ 
\kp\sqrt{\fl{\gm_z}{2\pi\vep}}= 
\dt \vep^{4/2}\sqrt{\fl{\gm_z}{2\pi}}, &\qquad d=2, \\ 
\kp=\dt \vep^{5/2}, &\qquad d=3,\\ 
\ea\right. 
\ee 
and for a condensate with strong repulsive interactions 
\be 
\label{ufs} 
\kp_d:=\kp_d^s=\left\{\ba{ll} 
\fl{\pi}{9}\left(\fl{15}{4\pi}\right)^{8/5}(\kp\gm_y\gm_z)^{3/5}= 
\fl{(\dt \gm_y \gm_z)^{3/5} \vep^{3/2} \pi }{9}\; 
 \left(\fl{15}{4\pi}\right)^{8/5}, &\qquad d=1, \\ 
\fl{5}{7}\left(\fl{4\pi}{15}\right)^{1/5}\fl{(\kp\gm_z)^{4/5}}{\gm_y^{1/5}}= 
(\dt\gm_z)^{4/5} \vep^{4/2}\left(\fl{4\pi}{15  \gm_y} 
\right)^{1/5}\; \fl{5}{7}, &\qquad d=2, \\ 
\kp=\dt \vep^{5/2}, &\qquad d=3.\\ 
\ea\right. 
\ee 
 
We call a $d$-dimensional ($d=1$ or $2$) condensate 
`very weakly interacting', if $\vep=O(1)$ and 
$|\kp_d|=|\kp_d^w|\ll1$, which implies $|\dt|\sqrt{\gm_z}\ll1$ in 2d and 
$|\dt|\sqrt{\gm_y\gm_z}\ll1$ after reduction to 1d.

\bigskip 
 
\begin{remark} 
 By using very elongated trapping potentials it is now possible to produce 
weakly interacting condensates that are `truly' in the 1D regime 
and the assumption that the condensate wave function factorizes is 
fulfilled with high accuracy. In cases where the interactions 
cannot be neglected low dimensional simulations 
\cite{LDsimulations} should be viewed as model calculations where 
the effective interaction strength in the reduced equation is 
estimated by Eq.~(\ref{ufw}) for a weakly interacting condensate 
or (\ref{ufs}) for a condensate with strong repulsive 
interactions. More detailed studies go beyond the GPE to describe 
low dimensional BEC's \cite{BeyondGPE}. 
\end{remark}

\subsection{Geometrical optics $\vep \to 0$} 
 
We set 
\[\psi(\bx,t)=\sqrt{n(\bx,t)}\exp\left(\fl{i}{\vep}S(\bx,t)\right), \] 
where $n=|\psi|^2$ and $S$ is the phase of the wave-function. 
Inserting into the GP-equation (\ref{gpe2}) and 
separating real and imaginary parts give 
 \bea 
&&n_t+{\rm div}\, (n\; \btd S)=0, \label{eqtrans}\\ 
&&S_t+\fl{1}{2}|\btd S|^2 +\kp 
n+\fl{1}{2}\left(x^2+\gm_y^2y^2+\gm_z^2z^2 
\right)=\fl{\vep^2}{2}\;\fl{1}{\sqrt{n}}\ \btu \sqrt{n}. 
\label{eqHJ} 
 \eea 
The equation (\ref{eqtrans}) is the transport equation for the 
atom density and (\ref{eqHJ}) the Hamilton-Jacobi equation for 
the phase. 
 
 By formally passing to the limit 
$\vep\to 0$ (cf. \cite{SAGardiner}), we obtain the system \bea \label{HJ1} 
&&n^0_t+{\rm div}\, (n^0\; \btd S^0)=0,\\ 
\label{HJ2} 
&&S^0_t+\fl{1}{2}|\btd S^0|^2 +\kp n^0+\fl{1}{2}\left(x^2+\gm_y^2y^2
+\gm_z^2z^2 
\right)=0. 
 \eea 
It is well known that this limit process is only correct in the 
defocusing case $\kp>0$ before caustic onset, i.e. in 
time-intervals where the solution of the Hamilton-Jacobin equation 
(\ref{eqHJ}) coupled with the atom-number conservation equation 
(\ref{eqtrans}) is smooth. After the breakdown of regularity 
oscillations occur which make the term 
$\fl{\vep^2}{2}\;\fl{1}{\sqrt{n}}\ \btu \sqrt{n}$ at least $O(1)$ 
such that the validity of the formal limit process is destroyed. 
The limiting behavior after caustics 
appear is not understood yet except in the one-dimensional case 
without confinement, see \cite{Shen}. Also, the focusing case 
$\kappa < 0$ is not fully understood yet.

\section{Numerical approximation}\label{sna} 
\setcounter{equation}{0} 
 
In this section we present a time-splitting Fourier spectral 
method, which was used by Bao et al.~to numerically solve the 
Schr\"{o}dinger equation in the semiclassical regime 
\cite{Bao1,Bao2}. We reiterate that neither time splitting 
discretisations nor Fourier spectral
methods are new, both have been applied successfully to
many PDE problems \cite{Canuto,spectral,Strang}. 
Here we adapt the combination of both
techniques to the GP equation and infer computational domain and
mesh size controls from analytical (perturbation) results.
The merit of this method is that it is 
unconditionally stable, time reversible, time-transverse 
invariant, and conserves the total particle number. Also, it has 
very favorable properties with respect to efficiently choosing the 
spatial/temporal grid in dependence of the semiclassical parameter 
$\vep$. For simplicity of notation  we shall introduce the method 
in one space dimension $(d=1)$. Generalizations to $d>1$ are 
straightforward for tensor product grids and the results remain 
valid without modifications. For $d=1$, the equation (\ref{gpeg}) 
with periodic boundary conditions becomes \bea \label{sdge1d} 
&&i\vep \pl{\psi(x,t)}{t}=-\fl{\vep^2}{2}\psi_{xx}(x,t)+ 
\fl{x^2}{2}\psi(x,t) 
+ \kp_1 |\psi(x,t)|^2\psi(x,t), \ a<x<b,\qquad \\ 
\label{sdgi1d} 
&&\psi(x,t=0)=\psi^0(x), \qquad  a\le x\le b, \\ 
\label{sdgb1d} 
&&\psi(a,t)=\psi(b,t),\qquad  \psi_x(a,t)=\psi_x(b,t), 
\qquad t>0.\qquad 
\eea 
 
We choose the spatial mesh size $h=\btu x>0$ with $h=(b-a)/M$ for $M$ an 
even positive integer, the time step $k=\btu t>0$ 
and let  the grid points and the time step be 
\[ x_j:=a+j\;h, \qquad t_n := n\; k, \qquad j=0,1,\cdots, M, \qquad 
n=0,1,2,\cdots   \] 
Let $\psi^n_j$ be the approximation of $\psi(x_j,t_n)$ and 
$\psi^n$ be the solution vector with components $\psi_j^n$. 
 
\medskip 
 
{\bf Time-splitting spectral method (TSSP)}. From time $t=t_n$ to 
$t=t_{n+1}$,  the GPE (\ref{sdge1d}) is solved in two splitting 
steps. One solves first \be \label{fstep} i\vep \psi_t=- 
\fl{\vep^2}{2} \psi_{xx}, \ee for the time step of length $k$, 
followed by solving \be \label{sstep} i\vep 
\pl{\psi(x,t)}{t}=\fl{x^2}{2}\psi(x,t) + \kp_1 
|\psi(x,t)|^2\psi(x,t), \ee for the same time step. Equation 
(\ref{fstep}) will be discretized in space by the Fourier spectral 
method and integrated in time {\it exactly}. For 
$t\in[t_n,t_{n+1}]$, the ODE (\ref{sstep}) leaves $|\psi|$ 
invariant in $t$ \cite{Bao1,Bao2} and  therefore  becomes \be 
\label{sstepp} i\vep \pl{\psi(x,t)}{t}=\fl{x^2}{2}\psi(x,t) + 
\kp_1 |\psi(x,t_n)|^2\psi(x,t) \ee and thus can be integrated {\it 
exactly}. From time $t=t_n$ to $t=t_{n+1}$, we combine the 
splitting steps via the standard Strang splitting: \bea 
&&\psi^*_j=e^{-i(x_j^2/2+\kp_1 |\psi_j^n|^2)k/(2\vep)}\;\psi_j^n,  \nn\\ 
&&\psi_j^{**}=\fl{1}{M}\sum_{l=-M/2}^{M/2-1} 
  e^{-i\vep k\mu_l^2/2}\;\widehat{\psi}^*_l\;e^{i\mu_l(x_j-a)}, 
    \qquad j=0,1,2,\cdots,M-1,\nn\\ 
\label{schm23} 
&&\psi^{n+1}_j=e^{-i(x_j^2/2+\kp_1 |\psi_j^{**}|^2)k/(2\vep)}\;\psi_j^{**}, 
 \qquad j=0,1,2,\cdots,M-1, 
\eea 
where $\widehat{\psi}^*_l$, the Fourier 
coefficients of $\psi^*$, are defined as 
\be 
\label{Fouv1} 
\mu_l=\fl{2\pi l}{b-a},\quad \widehat{\psi}^*_l=\sum_{j=0}^{M-1} 
 \psi^*_j\;e^{-i\mu_l(x_j-a)},  \quad l=-\fl{M}{2},\cdots,\fl{M}{2}-1. 
\ee 
The overall time discretization error comes solely from the splitting, 
which is second order in $k$ for fixed $\vep>0$. 
The spatial discretization is of spectral (i.e. `infinite') order 
of accuracy for $\vep>0$ fixed. An error analysis for linear 
Schr\"{o}dinger equations taking into account the $\vep$-dependence 
of the global error can be found in \cite{Bao1} (where it is 
shown that $k=O(1)$ and $h=\fl{b-a}{M}=O(\vep)$ give 
correct observable), numerical tests for 
nonlinear problems in the semiclassical regimes in \cite{Bao2}. 
More restrictive meshing strategies are typically necessary in nonlinear 
cases, cf. Section 4.1. 
\bigskip 
 
 For comparison purposes we review now alternative numerical methods 
\cite{Adh1,Tosi2,Edwards} which are 
currently used for solving the 
Gross-Pitaevskii equation of BEC. One is the Crank-Nicolson 
finite difference (CNFD) scheme \cite{Adh1}: 
\beas 
\label{cnfd1} 
&&\fl{\psi_j^{n+1}-\psi_j^n}{k}=\fl{i\vep}{4h^2}\left[\psi_{j+1}^{n+1} 
-2\psi_j^{n+1}+\psi_{j-1}^{n+1}+\psi_{j+1}^{n} 
-2\psi_j^{n}+\psi_{j-1}^{n}\right]\nn\\ 
&&\qquad \qquad -\fl{i\;x_j^2}{4\vep}(\psi_j^{n+1} 
+\psi_j^n)-\fl{i\kp_1}{2\vep}|\psi_j^n|^2(\psi_j^{n+1} 
+\psi_j^n), \quad  j=1,2,\cdots, M,\\ 
&&\psi_0^{n+1}=\psi_M^{n+1}, \qquad \psi_{M+1}^{n+1}=\psi_1^{n+1},\\ 
\label{cnfd2} 
&&\psi_j^0=\psi_0(x_j), \qquad j=0,1,\cdots, M. 
\eeas 
Another one is the Crank-Nicolson spectral (CNSP) method: 
\beas 
\label{cnfd3} 
&&\fl{\psi_j^{n+1}-\psi_j^n}{k}=\fl{i\vep}{4} 
\left[\left.D_{xx}^f\psi^{n+1}\right|_{x=x_j}+ 
\left.D_{xx}^f\psi^{n}\right|_{x=x_j}\right] 
-\fl{i\;x_j^2}{4\vep}(\psi_j^{n+1}-\psi_j^n)\nn\\ 
&&\qquad \qquad -\fl{i\kp_1}{2\vep}|\psi_j^n|^2(\psi_j^{n+1}+\psi_j^n),\\ 
&&\psi_j^0=\psi_0(x_j), \qquad j=0,1,\cdots, M; 
\eeas 
where $D_{xx}^f$, a spectral differential operator 
approximation of $\partial_{xx}$, is defined as 
\be 
\label{Dxx} 
\left.D_{xx}^f U\right|_{x=x_j} 
=-  \sum_{l=-M/2}^{M/2-1} 
\mu_l^2 (\widehat{U})_l\; e^{i\mu_l(x_j-a)}. 
 \ee 
Both methods are unconditionally stable, time reversible, conserve 
the total particle number but they are \underline{not} time 
transverse-invariant.  We do not present comparism tests with 
fully implicit and fully explicit finite
difference methods since they are not at all competitive with the time
splitting-spectral method. Generally 
1) they require severe stability constraints on the mesh sizes, 
2) they do not conserve the total particle number,
3) they are not time 
transverse invariant.
For a mathematical analysis of FD-methods for Schr\"{o}dinger 
type equations in
semiclassical regimes we refer to \cite{MaPiPo,MPPS}.


 
 

\section{Numerical examples}\label{sne} 
\setcounter{equation}{0} 
 
In this section, we first perform a numerical comparison of TSSP, 
CNFD and CNSP in terms of accuracy and mesh size strategy for a 1d 
defocusing GPE. Then we apply the TSSP for solving 1d, 2d and 3d 
GPEs of Bose-Einstein condensation. Furthermore we also give a 
physical discussion on our numerical results.

In our computations, the initial condition for (\ref{gpeg}) 
is always chosen in WKB form: 
\be 
\label{init} 
\psi(\bx,t=0)=\psi^0(\bx)=A^0({\bf x})\; 
e^{i S^0({\bf x})/\vep}, 
 \ee 
with $A^0$ and $S^0$ real valued, regular and with $A^0({\bf x})$ 
decaying to zero sufficiently fast as $|{\bf x}|\to\ift$. We 
compute with TSSP on a domain, which is large enough 
(as controlled by the initial data) such that the 
periodic boundary conditions do not introduce a significant 
aliasing error relative to the whole space problem. 
There are certainly more sophisticated analysis for 
controlling aliasing errors, however these do not 
significantly improve the results for exponentially 
decaying initial densities. 
 
To quantify the numerical results, we define the condensate 
widths along the $x$, $y$ and $z$-axis as 
\be 
\label{cw} 
\sg_x=\sqrt{\langle (x-\langle x\rangle)^2\rangle}, 
\qquad \sg_y=\sqrt{\langle (y-\langle y\rangle)^2\rangle}, 
\qquad \sg_z=\sqrt{\langle (z-\langle z\rangle)^2\rangle}, 
 \ee 
where brackets denote space averaging with respect to the position 
density 
 \[ 
\langle f \rangle \equiv \int_{{\Bbb R}^d}\; f(\bx) |\psi(\bx,t)|^2\; 
d\bx. 
\]

\subsection{Comparisons of different methods} 
\label{Compdiff} 
 
 {\bf Example 1} 1d defocusing condensate, 
i.e. we choose $d=1$ 
in (\ref{gpeg}) with positive $\kp_1$. 
The initial condition is taken as 
 \be 
\label{inite1} \psi(x,0)=\fl{1}{(\pi \vep)^{1/4}}e^{-x^2/(2 
\vep)}, \qquad x\in{\Bbb R}. 
 \ee 
We choose $\vep=0.1$ and $\kappa_1=1.2649$ and solve this 
problem on $[-16,16]$, i.e. $a=-16$ and $b=16$ with periodic 
boundary conditions. Let $\psi$ be the `exact' solution which is 
obtained numerically by using TSSP with a very fine mesh and time 
step, e.g., $h=\fl{1}{256}$ and $k=0.00001$, and $\psi_{h,k}$ be 
the numerical solution obtained by using a method with mesh size 
$h$ and time step $k$. 
 
\bigskip 
 
First we compare the discretization error in space. We choose a 
very small time step, e.g., $k=0.00002$ such that the error from 
the time discretization is negligible compared to the spatial 
discretization error, and solve the GPE using different methods 
and varying spatial mesh sizes $h$. Table 1 lists the numerical 
errors $\|\psi(t)-\psi_{h,k}(t)\|_{l^2}$ at $t=2$ for varying 
spatial mesh sizes $h$. Clearly TSSP and CNSP show roughly the 
same errors due to the fact that the temporal discretization is 
almost `exact'. 
 
\begin{table}[htbp] 
\begin{center} 
\begin{tabular}{ccccccc}\hline 
 Mesh   &$h=\fl{1}{4}$ &$h=\frac{1}{8}$ &$h=\frac{1}{16}$ 
   &$h=\frac{1}{32}$ &CPU time\\ 
\hline 
TSSP &$0.2248$ &$2.048E-2$ &$3.641E-5$ &$7.982E-10$ &0.01s\\ 
CNSP &$0.2248$ &$2.048E-2$ &$3.642E-5$ &$8.538E-8$ &15.54s\\ 
CNFD &$0.6314$ &$0.3380$ &$8.784E-2$ &$2.801E-2$ &1.27s\\ 
\hline 
\end{tabular} 
\end{center} 
Table 1: Spatial discretization error analysis: 
$\|\psi(t)-\psi_{h,k}(t)\|_{l^2}$ 
at time $t=2$ under $k=0.00002$. The CPU time is counted at the same 
accuracy (i.e. $\|\psi(2)-\psi_{h,k}(2)\|_{l^2}\approx 3.65E-5$) and
 on an AlphaServer DS20  workstation. For that accuracy, 
TSSP needs $h=\fl{1}{16}$ and $k=0.001$, CNSP needs $h=\fl{1}{16}$ and
$k=0.00002$ and CNFD needs $h=\fl{1}{1024}$ and $k=0.00001$
\end{table} 
 
Secondly, we test the discretization error in time. Again, we take 
$\vep=0.1$ and $\kp_1=1.2649$. Table 2 shows the numerical errors 
$\|\psi(t)-\psi_{h,k}(t)\|_{l^2}$ at $t=2$ with a very small mesh 
size $h=\fl{1}{32}$ for different time steps $k$ and different 
numerical methods. Here, CNSP and CNFD show almost no difference 
since the spatial discretization now is almost `exact'. 
 
\begin{table}[htbp] 
\begin{center} 
\begin{tabular}{cccccc}\hline 
 Time step   &$k=0.05$ &$k=0.025$ &$k=0.0125$ 
   &$k=0.00625$\\ 
\hline 
TSSP &$1.112E-2$ &$1.716E-3$ &$4.021E-4$ &$1.045E-4$\\ 
CNSP &$0.5215$ &$0.1247$ &$4.363E-2$ &$1.565E-2$ \\ 
CNFD &$0.5344$ &$0.13720$ &$6.121E-2$ &$3.723E-2$\\ 
\hline 
\end{tabular} 
\end{center} 
Table 2: Time discretization error analysis: 
$\|\psi(t)-\psi_{h,k}(t)\|_{l^2}$ 
at time $t=2$ under $h=\fl{1}{32}$. 
\end{table}

We also tested numerically the unconditional stability of the time-splitting 
spectral method, which was already proven rigorously in \cite{Bao1}. 
Numerical tests showed no significant accumulation of 
round-off errors and conservation of the discrete $l^2$-norm was 
observed up to $10$ significant digits for tests 
performed with $h=\fl{1}{32}$  and $k=0.2$, 
$k=0.05$, $k=0.01$ computing up to $t=4$.

%

At last, we test the $\vep$-resolution of different methods. Here we 
shall compare the meshing strategies required in order to get the 
`correct' condensate density $|\psi|^2$, for different methods 
when decreasing the semiclassical parameter $\vep$. Figure 1 shows 
the numerical results with different combinations of $\vep$, $h$, 
$k$ for different methods. Furthermore Figure 2 shows the 
evolution of $\rho=|\psi|^2$ in space-time and the 
condensate width as a function of time by using TSSP for 
$\vep=0.1$ and $\kp_1=1.2649$.

\bigskip 
 
  From Tables 1-3 and Figure 1, one can make the following observations: 
 
 (1). For TSSP, the spatial and temporal discretization errors are of 
spectral and second order accuracy, respectively. The admissible 
meshing strategy for obtaining the `correct' condensate density in 
the defocusing case is: $h=O(\vep)$ and $k=O(\vep)$. This method 
is explicit, unconditional stable 
 and its extension to 2d and 3d cases is straightforward 
without additional numerical difficulty. 
 
  (2). For CNSP, the spatial and temporal discretization errors are also 
of spectral and first order accuracy, respectively. But the 
admissible meshing strategy is: $h=O(\vep)$ and $k=o(\vep)$. 
Furthermore this method  is implicit and its extension to the 2d or 3d 
case is expensive except when an ADI technique is used, which 
destroys the spatial spectral accuracy. 
 
  (3). For CNFD, the spatial and temporal discretization errors are 
 of second and first order, respectively, 
 and the admissible meshing strategy is: 
$h=o(\vep)$ and $k=o(\vep)$ (see \cite{MaPiPo}). This method  is 
implicit and the remark of (2) applies. 
 
  Furthermore, the storage requirement of TSSP is less than
the other two methods.  The number of operations 
needed per time step is $O(M\ln M)$ for TSSP, at least 
$O(M^2)$ for CNSP, and $O(M)$ for CNFD when an ADI technique is used
in 2d and 3d, where $M$ is the total number of unknowns.
To attain the same order of accuracy, CNFD needs many more
grid points than TSSP.

\subsection{Applications} 
\label{secapp} 
 

 
 
 {\bf Example 2} 2d defocusing condensate,  i.e. we choose $d=2$ in 
 (\ref{gpeg}). 
We solve this problem on $[-8,8]^2$ with mesh size $h=\fl{1}{32}$ 
and time step $k=0.001$. We present computations for four cases: 
 
\bigskip 
 
\noindent {\it I. $O(1)$-interactions, zero initial phase data} 
\beas 
&&\vep=1.0, \qquad \gm_y=1.0,\qquad \kp_2=2.0\ (\gm_z=10.0,\quad 
\dt=1.586),\\ 
&&\psi(x,y,0)=\fl{1}{\sqrt{\pi\vep}}e^{-(x^2+y^2)/(2\vep)}. 
\eeas 
 
%
 
\smallskip 
 
\noindent {\it II. Very weak interactions, anisotropic condensate, 
nonzero initial phase } 
\beas 
&&\vep=1.0, \qquad \gm_y=2.0,\qquad \kp_2=0.1 \ (\gm_z=10.0,\quad 
\dt=0.0793),\\ 
&&\psi(x,y,0)=\fl{\gm_y^{1/4}}{\sqrt{\pi\vep}}e^{-(x^2+\gm_y y^2)/(2\vep)} 
e^{i\cosh\left(\sqrt{x^2+2y^2}\right)/\vep}. 
\eeas 
 
%
 
\smallskip 
 
\noindent {\it III. Strong interactions, nonzero initial phase 
data} \beas &&\vep=0.1, \quad \gm_y=1.0, \quad 
\mu_2^s=\sqrt{\kp_2\gm_y/\pi}, \quad \kp_2=1.2649\ 
(\gm_z=10.0,\quad 
\dt=65.5227),\\ 
&&\psi(x,y,0)=\left\{\ba{ll} \sqrt{(\mu_2^s-(x^2+y^2)/2)/\kp_2}\; 
e^{i\cosh\left(\sqrt{x^2+2y^2}\right)/\vep}, &\quad x^2+y^2< 2\mu_2^s,\\ 
0, &\quad \hbox{otherwise}. 
\ea\right. 
\eeas 
 


 
\smallskip

\noindent {\it IV. $O(1)$-interactions, anisotropic condensate with 
changing trap frequency } 
\beas 
&&\vep=1.0,  \quad \gm_y=2.0, \quad 
\vep_1=2.0,\quad \kp_2=2.0\ (\gm_z=10.0,\quad 
\dt=1.586),\\ 
&&\psi(x,y,0)=\fl{\gm_y^{1/4}}{\sqrt{\pi\vep_1}} 
e^{-(x^2+\gm_y y^2)/(2\vep_1)}. 
\eeas 
 
\bigskip 
 
 Figure 3 shows the surface plot of $\rho=|\psi|^2$ (labeled as $|u|^2$ in 
the figures) at time 
$t=40 $ and the condensate widths as a function of time for case I. 
Furthermore, Figure 4 shows similar results for case II, 
 Figure 5 for case III, 
 Figures 6-7  for case IV. 

\bigskip 
 
 {\bf Example 3} 2d focusing condensate,  i.e. we choose $d=2$ in 
in (\ref{gpeg}). 
We solve this problem on $[-10,10]^2$ with mesh size $h=\fl{1}{51.2}$ and 
time step $k=0.00005$. We present computations for three cases: 
 
\bigskip 
 
\noindent {\it I. $O(1)$-interactions, positive initial energy} 
\beas 
&&\vep=1.0,  \quad \gm_y=1.0, \quad 
 \kp_2=-2.0\ (\gm_z=10.0,\quad 
\dt=-1.586),\\ 
&&\psi(x,y,0)=\fl{1}{\sqrt{\pi\vep}}e^{-(x^2+y^2)/(2\vep)}. 
\eeas 
\smallskip 
 
\noindent {\it II. Strong interactions, negative initial energy} 
\beas 
&&\vep=0.3, \qquad \gm_y=1.0,\qquad \kp_2=-1.9718\ (\gm_z=10.0,\quad 
\dt=-7.545),\\ 
&&\psi(x,y,0)=\fl{1}{\sqrt{\pi\vep}}e^{-(x^2+y^2)/(2\vep)}. 
\eeas 
 
 
 
\bigskip 
 
  Figure 8 shows the surface plot of $\rho=|\psi|^2$ at time 
$t=40$ and the condensate widths as a function of time for case I. 
Furthermore, Figure 9 for case II. 

\bigskip 
 
 {\bf Example 4} 3d defocusing 
condensate,  i.e. we choose $d=3$ in 
in (\ref{gpeg}). 
We solve for $\bx\in [-8,8]^3$ with mesh size $h=\fl{1}{8}$ and 
time step $k=0.001$ and present computations for two cases: 
 
\bigskip 
 
\noindent {\it I. Anisotropic condensate with changing trap frequency} 
\beas 
&&\vep=1.0, \quad \gm_y=2.0, \quad \gm_z=4.0, 
\quad \vep_1=\fl{1}{4},\quad \kp_3=0.1 \ (\dt=0.1),\\ 
&&\psi(x,y,z,0)=\fl{(\gm_y\gm_z)^{1/4}} 
{\sqrt{(\pi\vep_1)^{3/4}}}e^{-(x^2+\gm_y y^2+\gm_z z^2)/(2\vep_1)}. 
\eeas 
 
\smallskip 
 
\noindent {\it II. Cylindrically symmetric condensate with changing trap 
frequency} 
\beas 
&&\vep=1.0, \quad \gm_y=1.0, \quad \gm_z=2.0, 
\quad \vep_1=\fl{1}{4}, \quad \kp_3=1.0\ (\dt=1.0), \\ 
&&\psi(x,y,z,0)=\fl{(\gm_y\gm_z)^{1/4}} 
{\sqrt{(\pi\vep_1)^{3/4}}}e^{-(x^2+\gm_y y^2+\gm_z z^2)/(2\vep_1)}. 
\eeas 
 
\bigskip 
 
Figure 10 shows  the condensate widths as a function of time for cases I and 
II. 
 
\bigskip 

   {\bf Example 5} $2$-dim  vortices in BEC. We choose $d=2$  and 
simulate the effect of stirring the (stable) condensate by adding 
a narrow, circularly moving Gaussian potential $W(\bx,t)$ to 
the stationary trap potential in (\ref{gpeg}). $W(\bx,t)$ represents, 
for example, a far-blue-detuned laser \cite{CD1}. We set 
\[W(\bx,t)= W_s(t) \exp\left[-4|\bx-\bx_s(t)|^2/V_s^2\right],\] 
with the center  $\bx_s(t)=(r_0\cos \og_s t, r_0\sin \og_s t)$ 
moving on a circle with radius $r_0$ and frequency $\og_s$. 
We start the simulation with the ground state in 2 dimensions 
(no stirrer at $t=0$) and minimize  transient 
effects by increasing the stirrer amplitude $W_s(t)$ linearly 
from $0$ at $t=0$ to a final value $W_s(t=\pi)=:W_f$ 
at $t=\pi$. The stirrer is then linearly withdrawn from 
$t=4\pi$ to $t=5\pi$  (after constant stirring, i.e. $W_s(t)=W_f$ 
for $\pi\le t\le 4\pi$) and the condensate is left to evolve freely after 
$t=5\pi$. We recall that $2$-dim  and $3$-dim 
vortices simulations were already performed in \cite{CD1,CD2,Du}, 
here we present this example in order to put 
our numerical method to an important physical test. 
 
We take the numerical values $\vep=1/\sqrt{50}$, $\kp_d=1$, 
$\gm_y=1$, $W_f=\sqrt{2}$, $V_s=1/50^{1/4}$, 
$r_0=2/50^{1/4}$ and $\og_s=1$ for our simulation. 
We remark that a vortex in the condensate is a point $\bx_w$ 
with $\psi(\bx_w)=0$ \underline{and} singular 
or undefined phase. 
 
In Figure 11, we show the contour plot of the density 
$|\psi(\bx,t)|^2$ at $t=12\pi$ (where the fluid has already 
settled down after the stirring) and $x,y$-sectional plots 
of the vortex centered 
at $(x\approx -0.141,y\approx-0.229)$. 
in fact  three vortices (labeled by 'X'), 
located at $(-0.141,-0.229)$, $(1.093,-0.0353)$ 
and $(0.282,1.481)$ were identified. 
For an analysis of vortex-formation 
in semiclassical limits 
of the Schr\"{o}dinger equation we refer to 
\cite{LFH}.

 
 
 
\bigskip

\subsection{Discussion} 
 
In Sec.~\ref{Compdiff} we compared different numerical methods for 
solving the GPE with the TSSP. Now we complete our investigation 
on the validity of using the TSSP for solving the GPE by comparing 
the results obtained in Sec.~\ref{secapp} with well known 
properties of Bose-Einstein condensates at very low temperatures. 
 
In example 1 we present 1d simulations. Initially the 
condensate is assumed to be in its noninteracting ground state 
when at $t=0$ repulsive interaction is 
turned on. In current experiments a change in the interaction 
strength can be achieved by applying external magnetic fields. 
Close to a Feshbach resonance the interaction strength shows a 
strong dependence on the magnetic field and even its sign can be 
changed by an appropriate choice of the magnetic field 
\cite{Donley}. At the same time we also change the trap potential 
by setting $\vep = 0.1$. These sudden changes lead to many rapid 
oscillations in the condensates (cf.~Fig.~2 a)). The 
dominant excitation caused by the interactions is, as expected 
\cite{collexp,colltheo}, the oscillation of the condensate width 
at approximately twice the trap frequency (cf.~Fig.~2 b)). 
 
Example 2 presents 2d simulations for various cases. In cases I 
there are $O(1)$ - interactions and in case II we investigate a 
weakly interacting condensate. We assume the condensate to be 
initially in the noninteracting ground state with possibly a 
nonuniform phase. As in example 1 turning on the interactions 
causes the condensate to oscillate at approximately twice the 
corresponding trap frequency. Higher excitations like in Fig.~2 a) 
 are not visible in Figs.~3 a), 4 a) since compared 
to example 1 the strength of the interactions is much smaller. A 
nonuniform phase like in II causes the amplitude of the 
oscillations to be different in $x$ and $y$ direction. Note also 
that while in case I the condensate immediately starts to expand 
(cf.~Fig.~3 b)) (due to the repulsive interactions) it starts to 
contract for the initial condition with a nonzero phase in case 
II. In case III we investigate the evolution of a strongly 
interacting condensate initially in the (approximate) Thomas Fermi 
ground state with an additional phase. Again, since this is not 
the exact ground state (see Remark 2.2) 
the width of the condensate starts to 
oscillate at about twice the trap frequency (cf.~Fig.~5 a)). In 
this case, however, due to the strong nonlinearity the 
oscillations along the $x$ and $y$ direction are coupled with each 
other as can be seen from Fig.~5 b). In case IV we investigate the 
effect of changing the trap frequency and turning on repulsive 
interactions (see Fig.~6). Like in the previous cases we find the 
dominant effect to be oscillations at about twice the 
corresponding trap frequency. The condensate initially starts to 
contract since the trap frequencies are increased at $t=0$. For 
the initial conditions chosen in this case the amplitudes are 
sufficiently large to swap the widths $\sigma_x$ and $\sigma_y$ 
(cf. Fig.~7), whereas for small changes in the trap frequency no 
swapping of the condensate widths would occur. The numerical 
results for the oscillations of a BEC obtained in example 2 agree 
very well with experimental and theoretical results 
\cite{collexp,colltheo}. 
 
In example 3 we show solutions for a focusing nonlinearity in 2d. 
Case I shows the effect of turning on $O(1)$ attractive 
interactions which leads to oscillations similar to those 
discussed in the previous examples (see Fig.~8). In case II a 
condensate with negative initial energy is shown. We have not 
discussed the reduction of the GPE to 2d for this case. Also, our 
simulations do not contain loss terms which become important in 
condensates at high densities. Thus we do not give a physical 
interpretation of these results. However, this example shows that 
the numerical method is applicable to the case of strong focusing 
nonlinearities in the GPE. Our numerical results confirm that the 
attractive GPE in 2d with negative initial energy will blow up at 
finite time (cf. Fig.~9). Furthermore, we point out that the TSSP 
allows the inclusion of loss terms into the GPE and it is also 
feasible to solve the GPE in 3d for the attractive case 
\cite{BaoNew}. Therefore the TSSP is a promising candidate for 
simulating the recent experiments on collapsing and exploding 
BEC's by Donley {\rm et al.} \cite{Donley2} which requires full 3d 
simulations and the inclusion of loss channels. 
 
Example 4 shows the effects of turning on repulsive interactions 
and changing the trap frequency in a 3d condensate. As expected 
from our previous simulations we see in Fig.~10 oscillations at 
twice the trap frequency in directions $x$, $y$ and $z$, 
respectively. The amplitude of the oscillations decreases with 
increasing frequency, i.e.~it becomes more difficult to excite 
oscillations for larger trap frequencies. This behavior is one of 
the basic assumptions allowing the reduction of the GPE to 2d and 
1d in the cases where one or two of the trap frequencies are much 
larger than the others (cf.~Sec.~2.3). 
 
The last example 5 shows the creation of vortices in a 2d BEC by 
stirring it with a blue detuned laser beam (see also 
\cite{CD1,CD2,vortexexp}) where we identify the creation of three 
vortices in the BEC as shown in Fig.~11. We note that since we 
study the GPE without taking into account the interaction between 
the condensate and a thermal cloud of atoms (additional 
dissipation) no stationary state showing an Abrikosov lattice of 
vortices is found as in recent experiments \cite{vortexexp} and 
numerical studies on the effects of a thermal cloud \cite{CD3} on 
the vortex formation. However, we point out that full 3d 
simulations including the effects of thermal particles with a very 
high precision are feasible based on the TSSP. 
 
Finally, we note that the TSSP is a very powerful versatile 
numerical method for solving the GPE which can be applied to a 
large number of different physical situations. The efficiency of 
this method and the high precision of the solutions make the TSSP 
a good choice for solving experimental situations that are 
numerically very demanding. Among these we believe that numerical 
studies of collapsing condensates with attractive interactions and 
multi-component condensates taking into account all experimentally 
relevant extensions of the GPE as well as on extensions of the GPE 
dealing with dissipation mechanisms \cite{Fudge} will be feasible 
by using the TSSP \cite{BaoNew}. 
 
\section{Summary}\label{sc} 
\setcounter{equation}{0} 
 
We studied a numerical method for solving the time-dependent GPE 
which describes trapped Bose-Einstein condensates at temperatures 
$T$ much smaller than the critical condensate temperature $T_c$. 
We started with the 3d GPE, scaled it to obtain a four-parameter 
model, and showed how to approximately reduce it to a 2d GPE and a 
1d GPE in certain limits. We provided the approximate ground state 
solution of the GPE in two extreme regimes: (very) weakly 
interacting condensates and condensates with strong repulsive 
interactions. Then, most importantly, we used the time-splitting 
spectral method in connection with analytical considerations based 
on perturbation theory (mesh-size control, dimension reduction) to 
solve the time-dependent GPE in 1d, 2d and 3d. Extensive numerical 
examples in 1d, 2d and 3d for weakly/strongly interacting 
condensates defocusing/focusing nonlinearity, and zero/nonzero 
initial phase data were presented to demonstrate the power of the 
time-splitting spectral numerical method. Finally, we want to 
point out that equations very similar to the GPE are also 
encountered in nonlinear optics. In the future we plan to apply 
this powerful numerical method to physically more complex systems 
like multi component condensates, collapsing condensates with 
attractive interactions and also to describe coherent atomic 
samples in wave guides.

\bigskip 
 
\begin{center} 
{\large \bf Acknowledgment} 
\end{center} 
 
W.B. acknowledges support  by the National University of Singapore 
 grant No. R-151-000-027-112. 
P.A.M. acknowledges support from 
the EU-funded TMR network `Asymptotic Methods in kinetic Theory', 
 from  his WITTGENSTEIN-AWARD 2000 
funded by the Austrian National Science Fund  FWF, and 
from Weizhu Bao's grant at NUS. 
D.J acknowledges support from the WITTGENSTEIN-AWARD of P. Zoller. 
This research was supported in part by the International Erwin 
Schr\"{o}dinger Institute in Vienna 
and the START project `Nonlinear Schr\"{o}dinger and quantum Boltzmann 
equations' (FWF Y137-TEC) of N.J. Mauser.  The three authors acknowledge 
discussion with N.J. Mauser and P. Zoller.

 
\newpage 
 
\begin{figure}[htb] 
\centerline{a)\psfig{figure=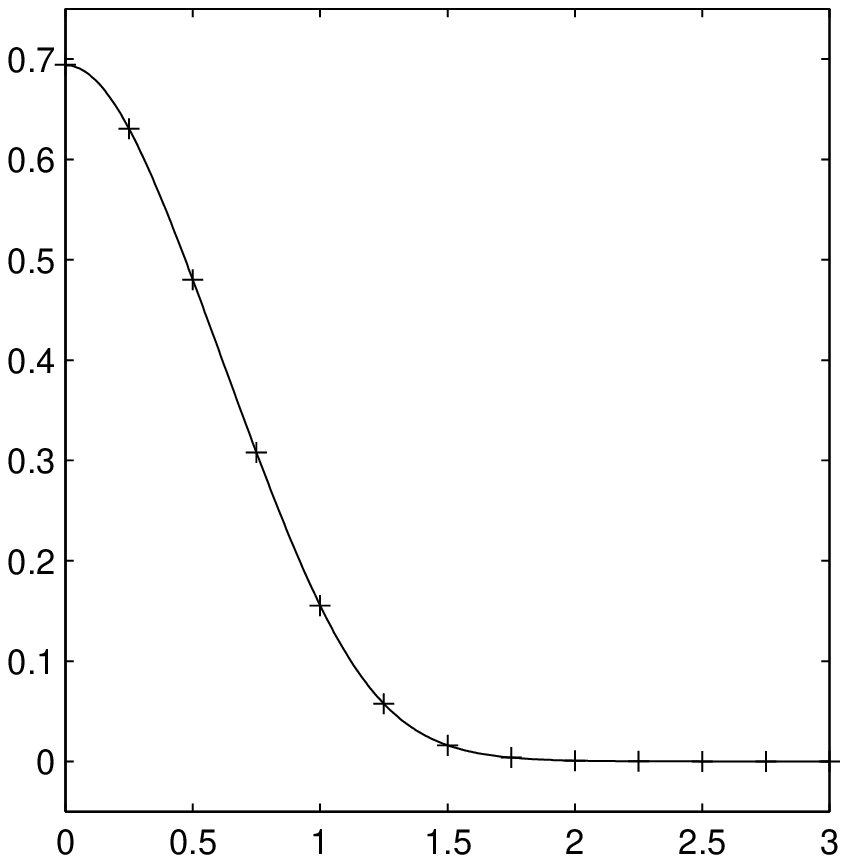,height=6cm,width=5cm,angle=0} 
\qquad \psfig{figure=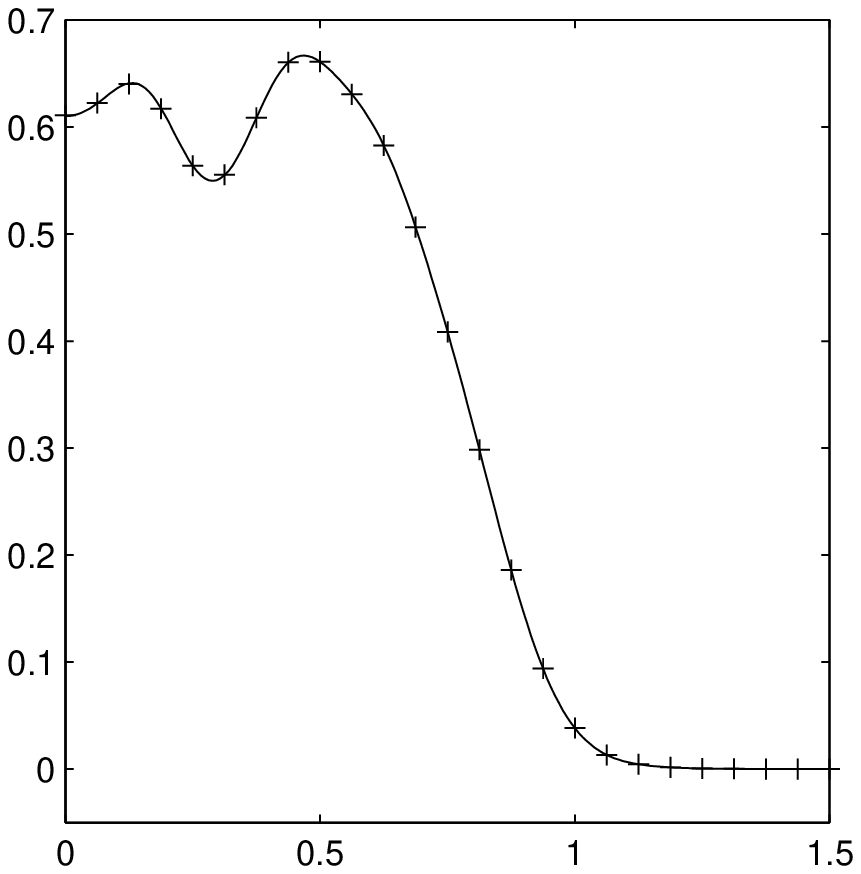,height=6cm,width=5cm,angle=0} 
\qquad \psfig{figure=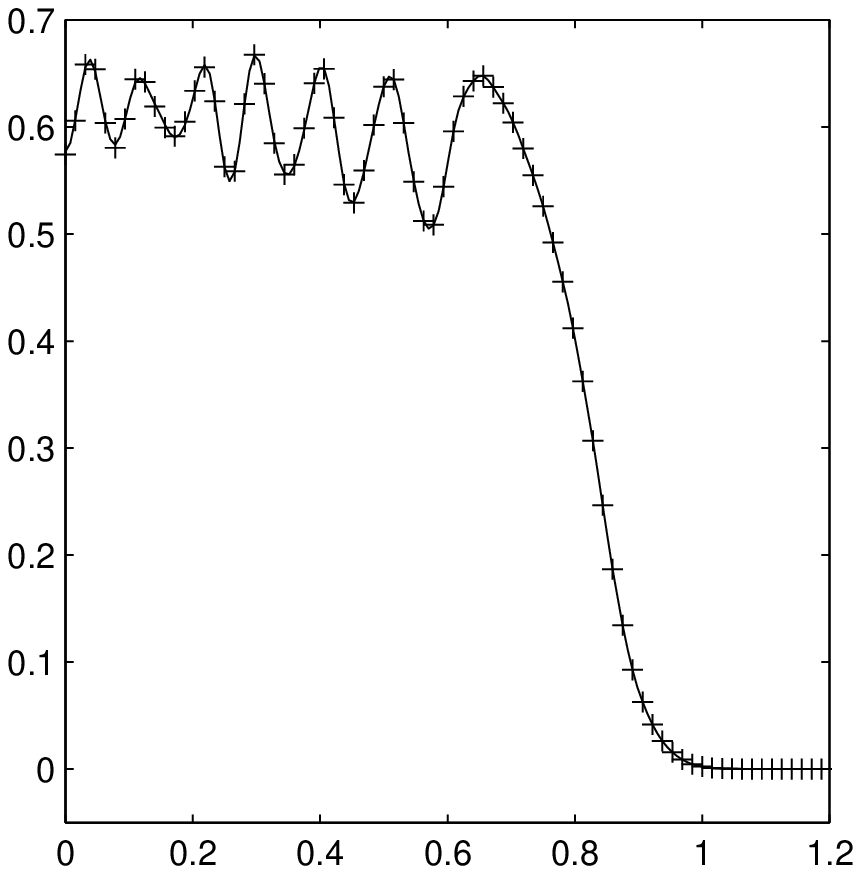,height=6cm,width=5cm,angle=0}} 
\centerline{b)\psfig{figure=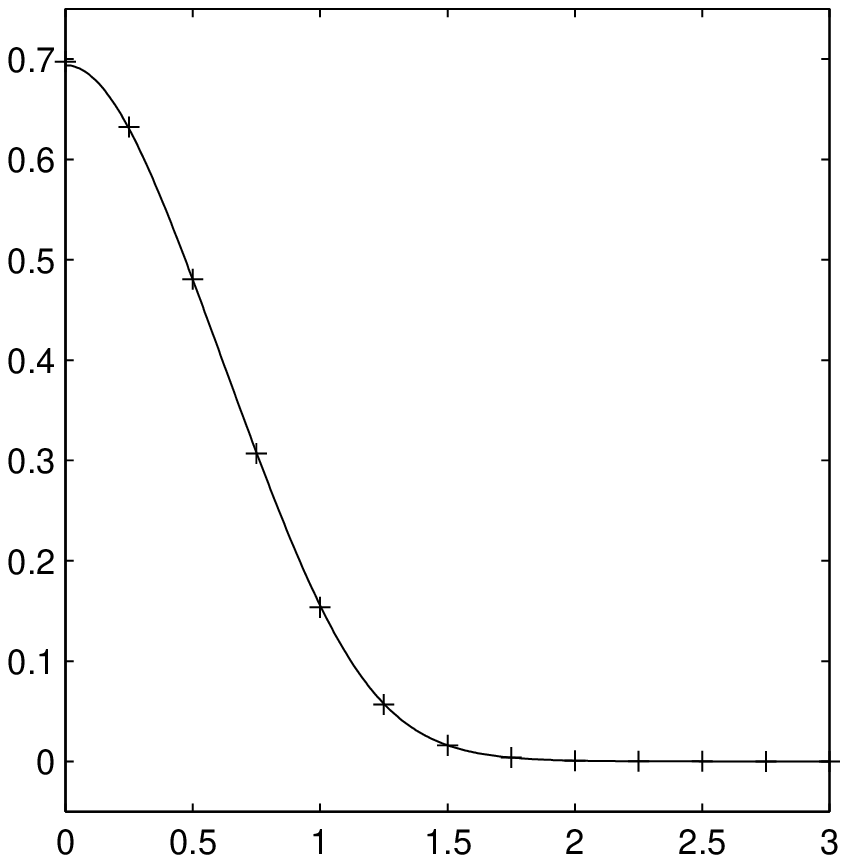,height=6cm,width=5cm,angle=0} 
\qquad \psfig{figure=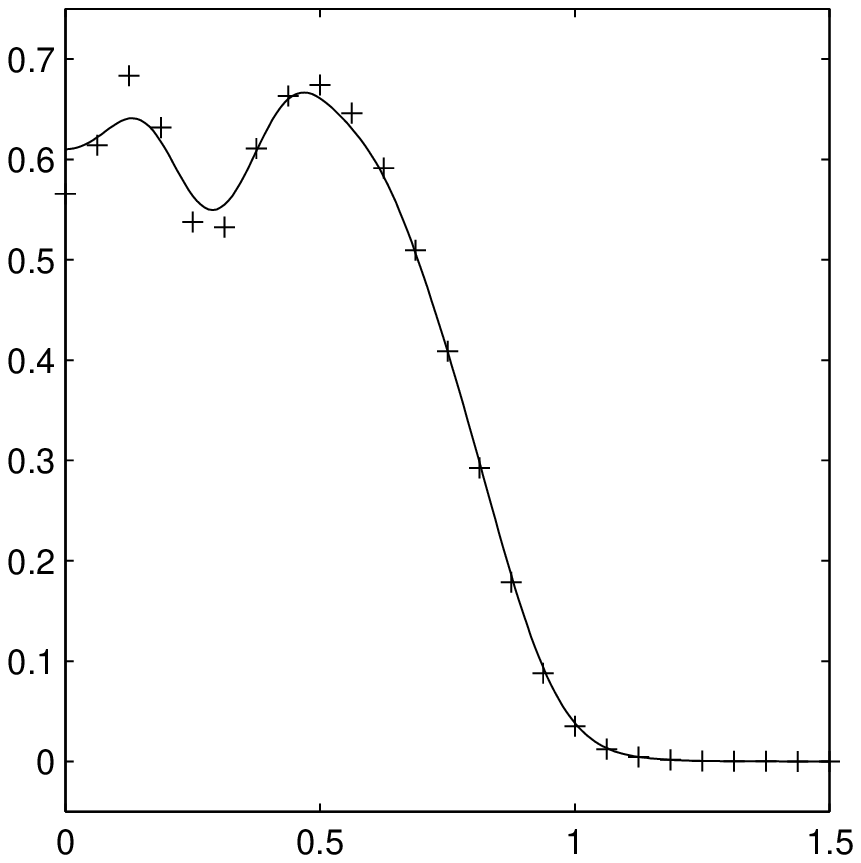,height=6cm,width=5cm,angle=0} 
\qquad \psfig{figure=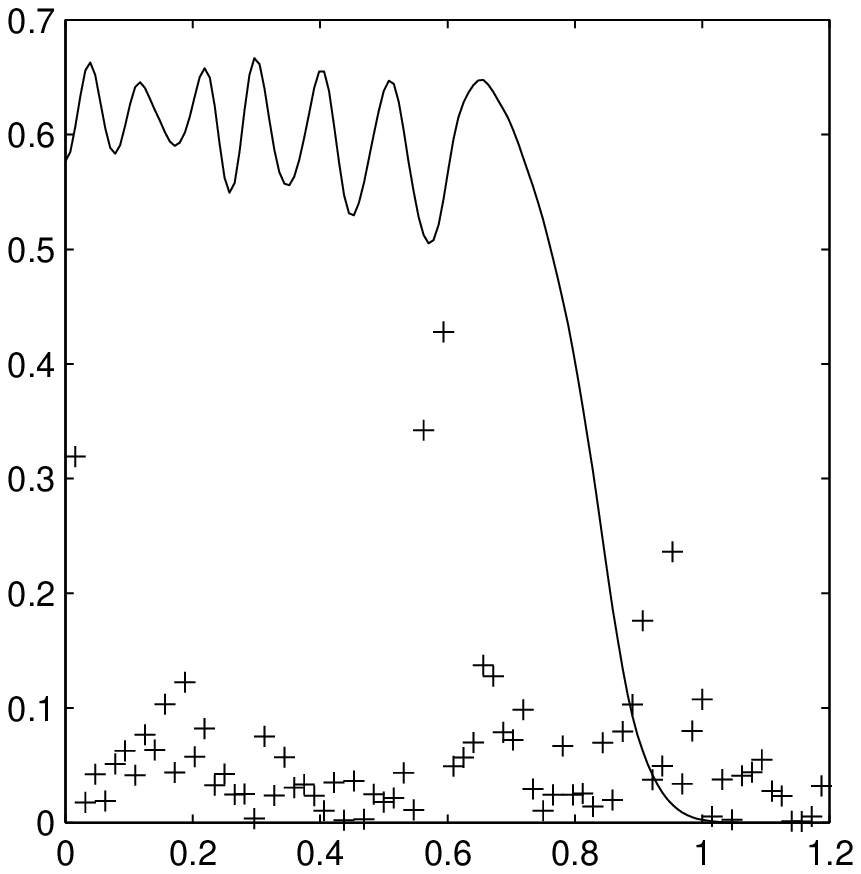,height=6cm,width=5cm,angle=0}} 
\centerline{c)\psfig{figure=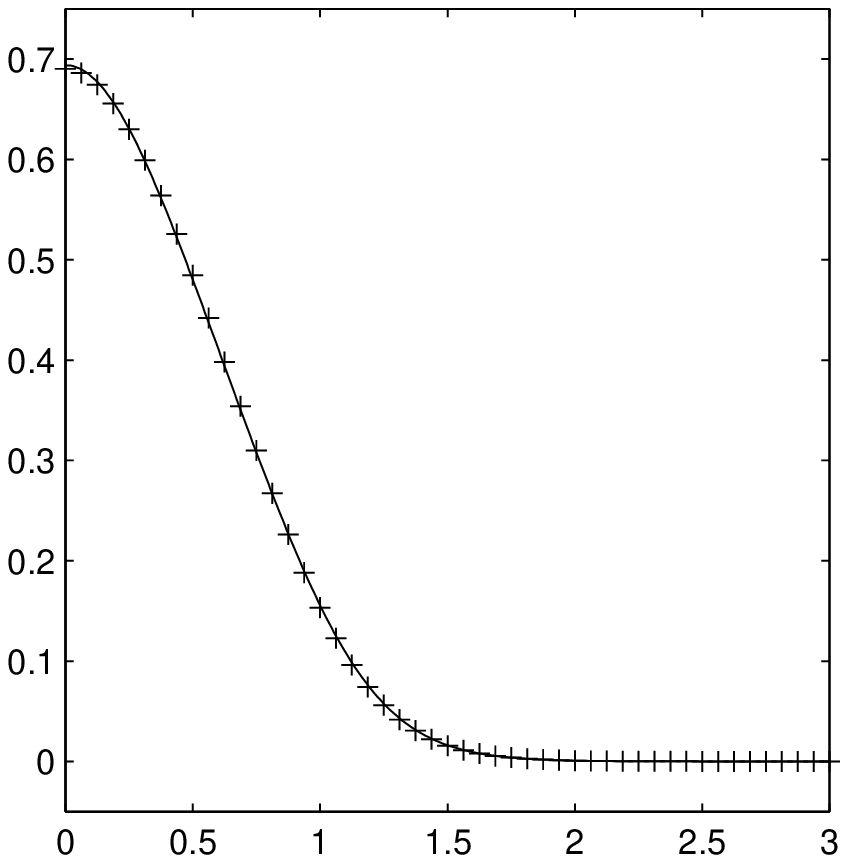,height=6cm,width=5cm,angle=0} 
\qquad \psfig{figure=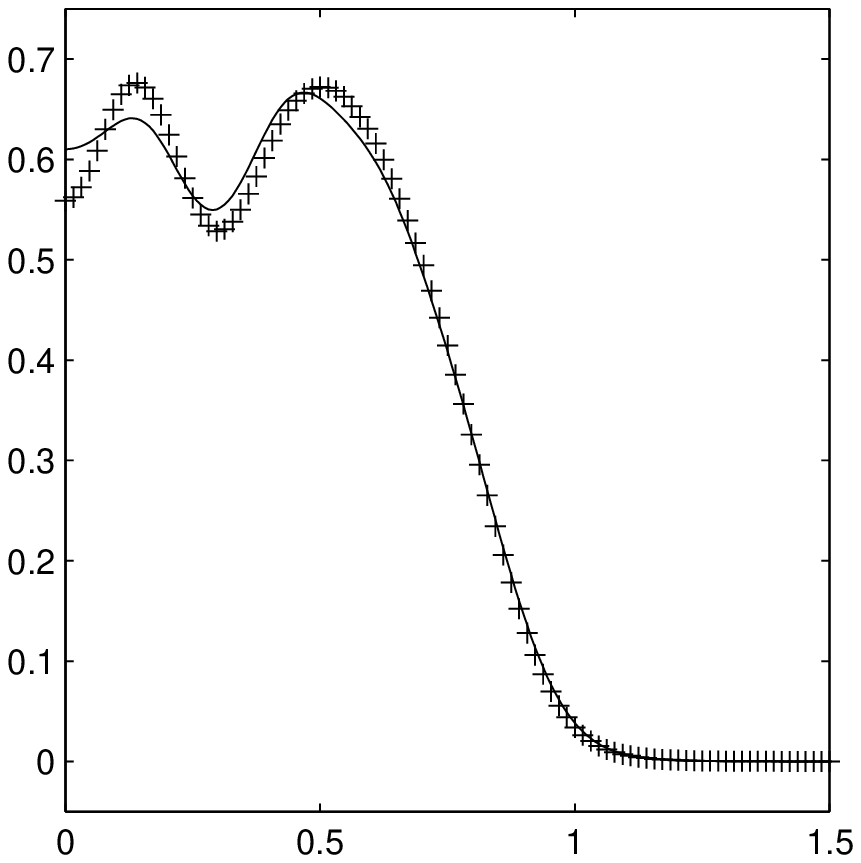,height=6cm,width=5cm,angle=0} 
\qquad \psfig{figure=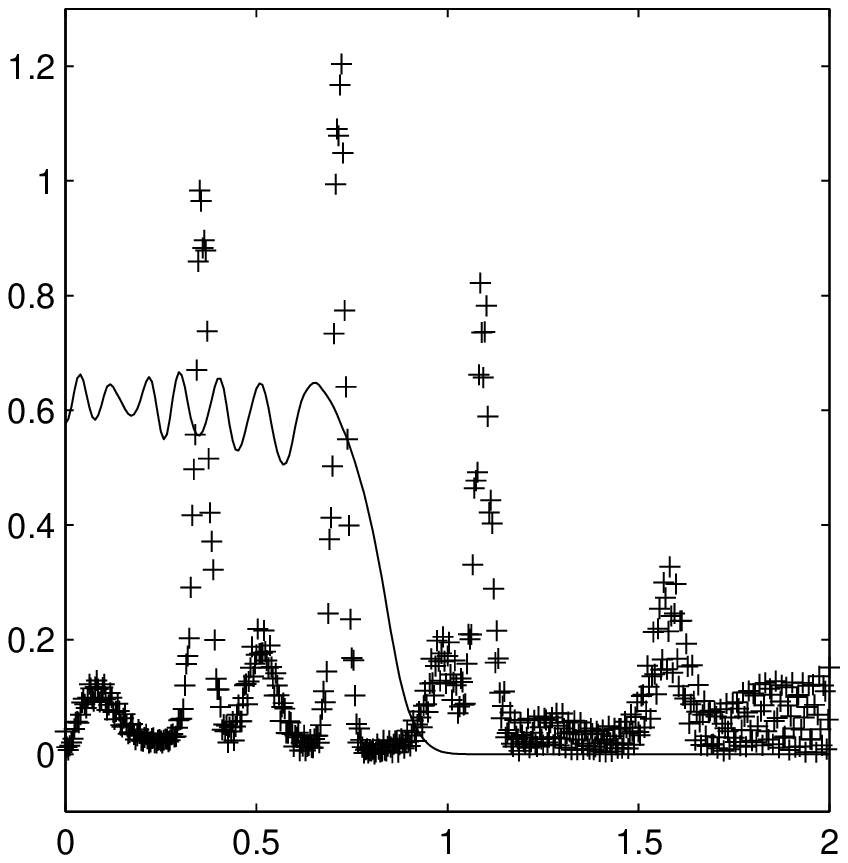,height=6cm,width=5cm,angle=0}}

 Figure 1:  $\vep$-resolution comparison 
in Example 1 for condensate density 
$|\psi|^2$ of different methods, 
`---': `exact' solution, `+ + +': numerical solution. 
a). TSSP: $\vep=0.4$, $h=\fl{1}{4}$, $k=0.04$ (left); 
 $\vep=0.1$, $h=\fl{1}{16}$, $k=0.01$ (middle); 
and $\vep=0.025$, $h=\fl{1}{64}$, $k=0.0025$ (right). 
b). CNSP:  $\vep=0.4$, $h=\fl{1}{4}$, $k=0.02$ (left); 
$\vep=0.1$, $h=\fl{1}{16}$, $k=0.005$ (middle); 
and $\vep=0.025$, $h=\fl{1}{64}$, $k=0.00125$ (right). 
c). CNFD: $\vep=0.4$, $h=\fl{1}{16}$, $k=0.02$ (left); 
$\vep=0.1$, $h=\fl{1}{64}$, $k=0.005$ (middle); 
and $\vep=0.025$, $h=\fl{1}{256}$, $k=0.00125$ (right). 
\end{figure}

 \begin{figure}[htb] 
\centerline{a)\psfig{figure=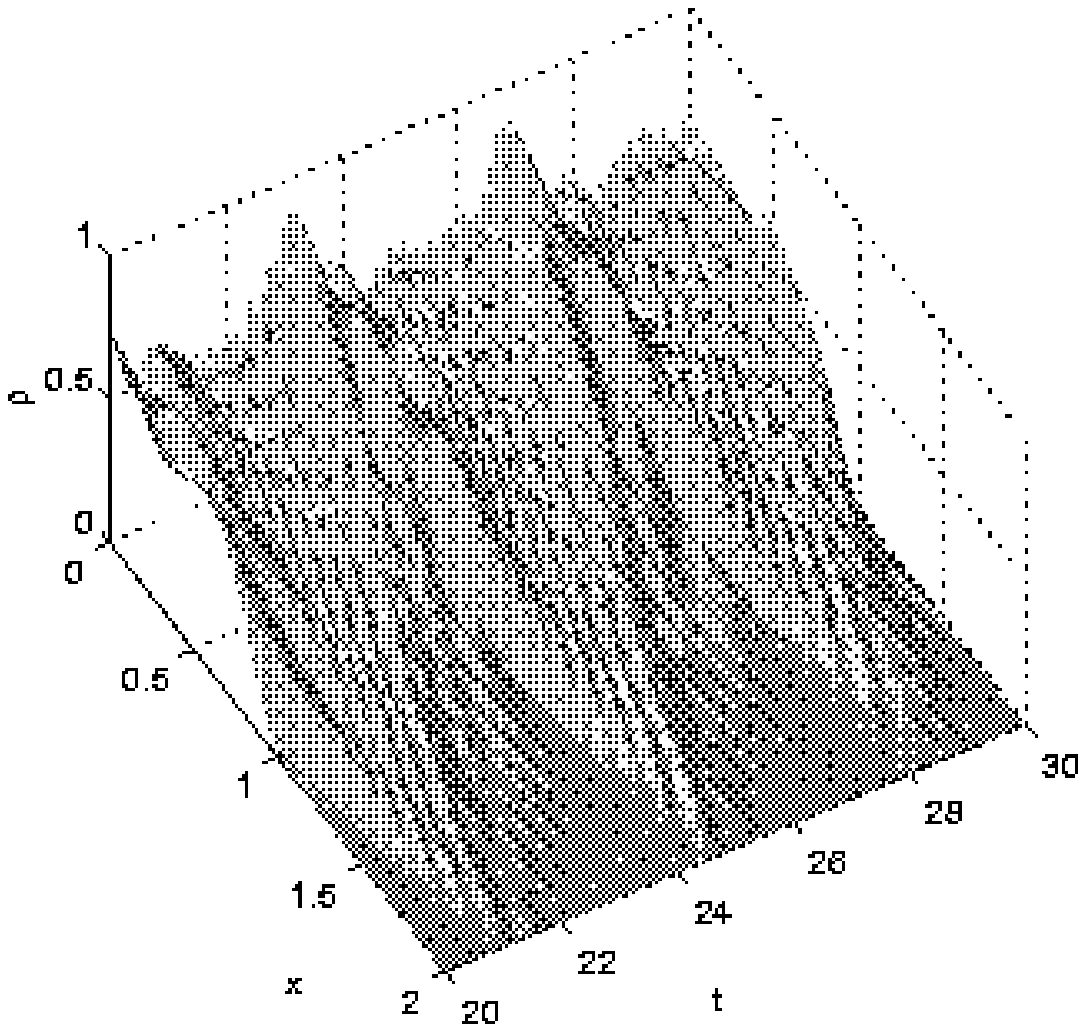,height=8cm,width=8cm,angle=0} 
b)\psfig{figure=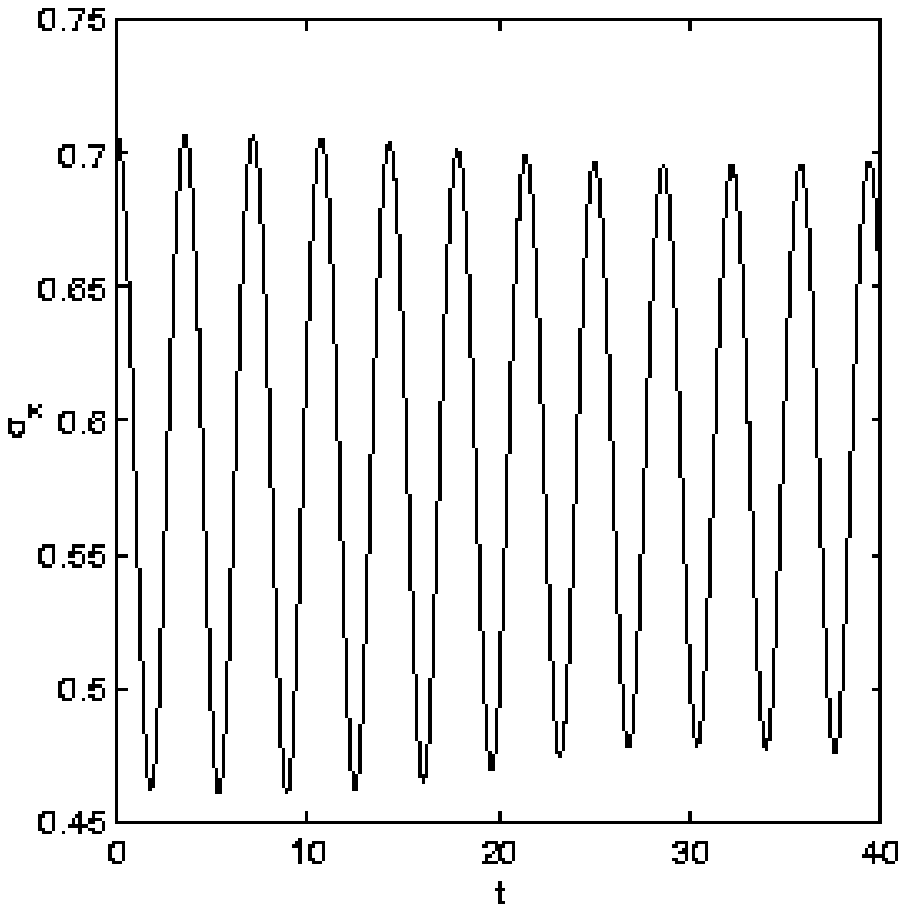,height=7cm,width=6cm,angle=0}} 
 
  Figure 2:  Numerical results in Example 1 
for $\vep=0.1$ and $\kp_1=1.2649$. \quad 
a). Evolution of the position density; 
b).  widths of the condensate as a function of time. 
 
 
\end{figure}

\clearpage

\newpage 
 
 \begin{figure}[htb] 
\centerline{a)\psfig{figure=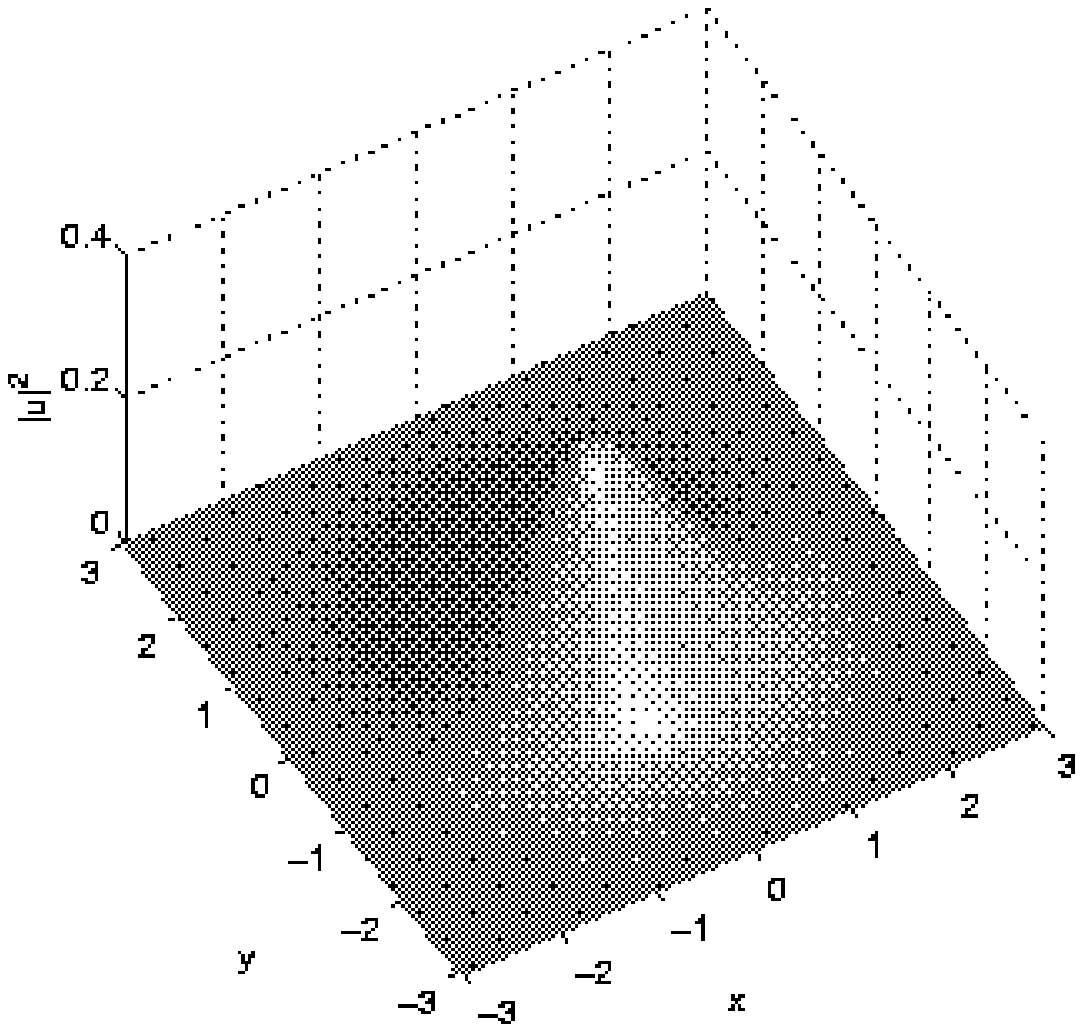,height=8cm,width=8cm,angle=0} 
b)\psfig{figure=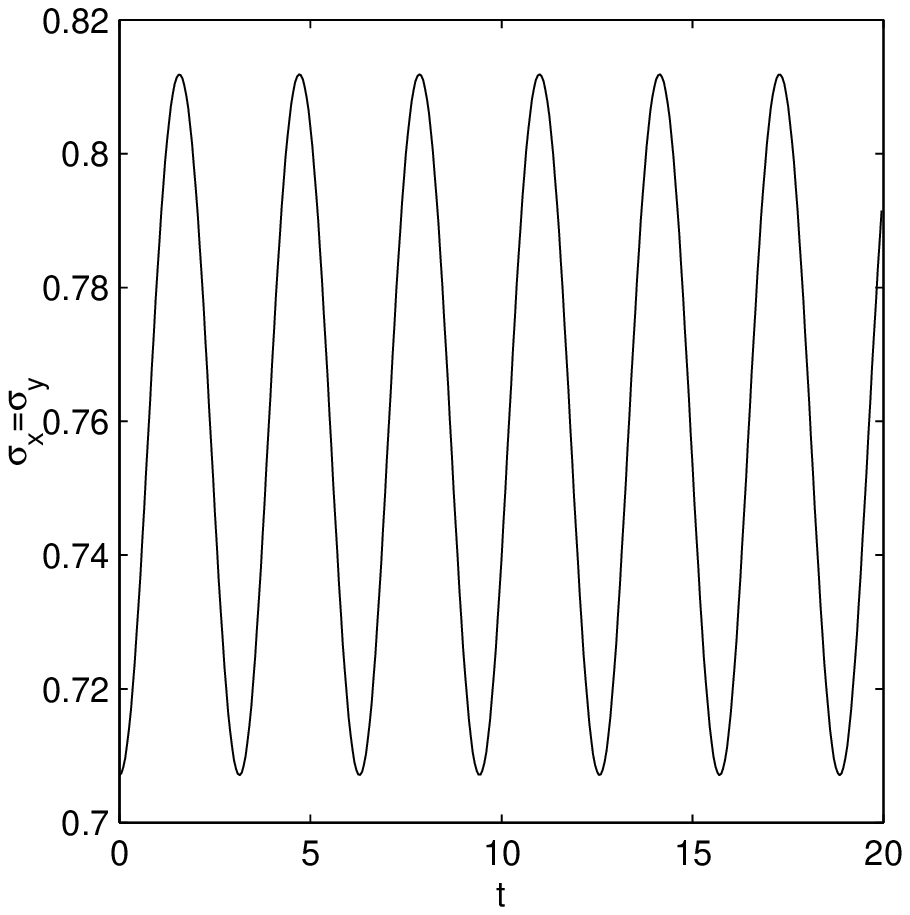,height=7cm,width=6cm,angle=0}} 
  Figure 3:  Numerical results in  Example 3 for case I. \quad 
a). Surface plot of the position density at $t=40$; 
b). widths of the condensate as a function of time. 
 
\vspace{2cm} 
 
 \centerline{a)\psfig{figure=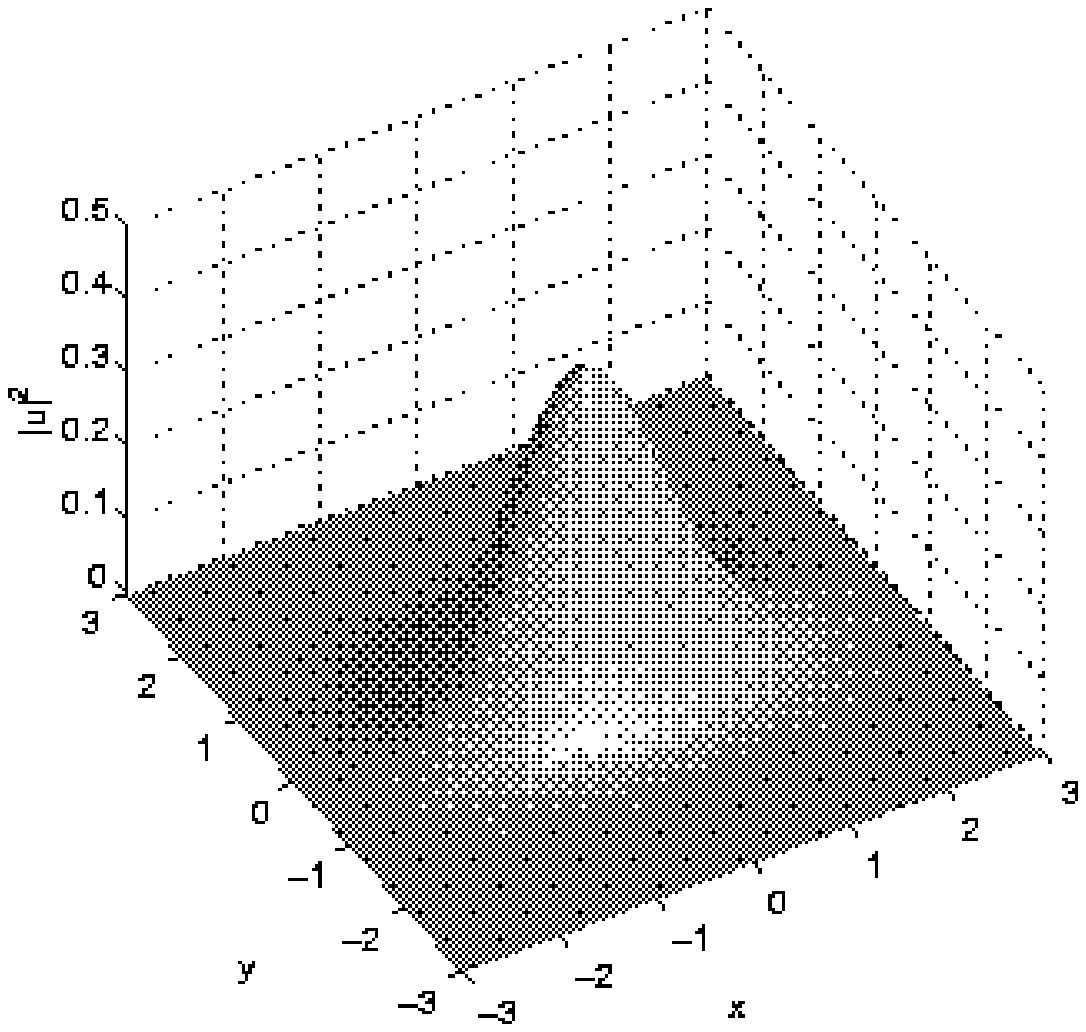,height=8cm,width=8cm,angle=0} 
b)\psfig{figure=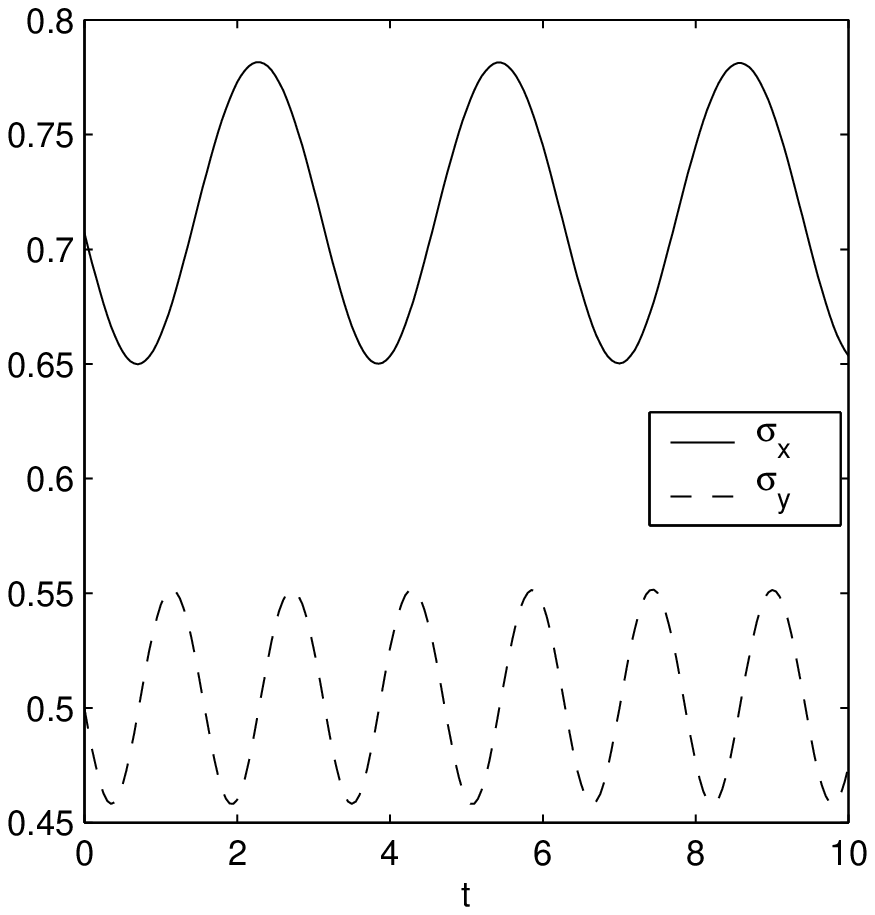,height=7cm,width=6cm,angle=0}} 
 
  Figure 4:  Numerical results in  Example 3 for case II. \quad 
a). Surface plot of the position density at $t=40$; 
b). widths of the condensate as a function of time. 
 
\end{figure}

\newpage 
\begin{figure}[htb] 
\centerline{a)\psfig{figure=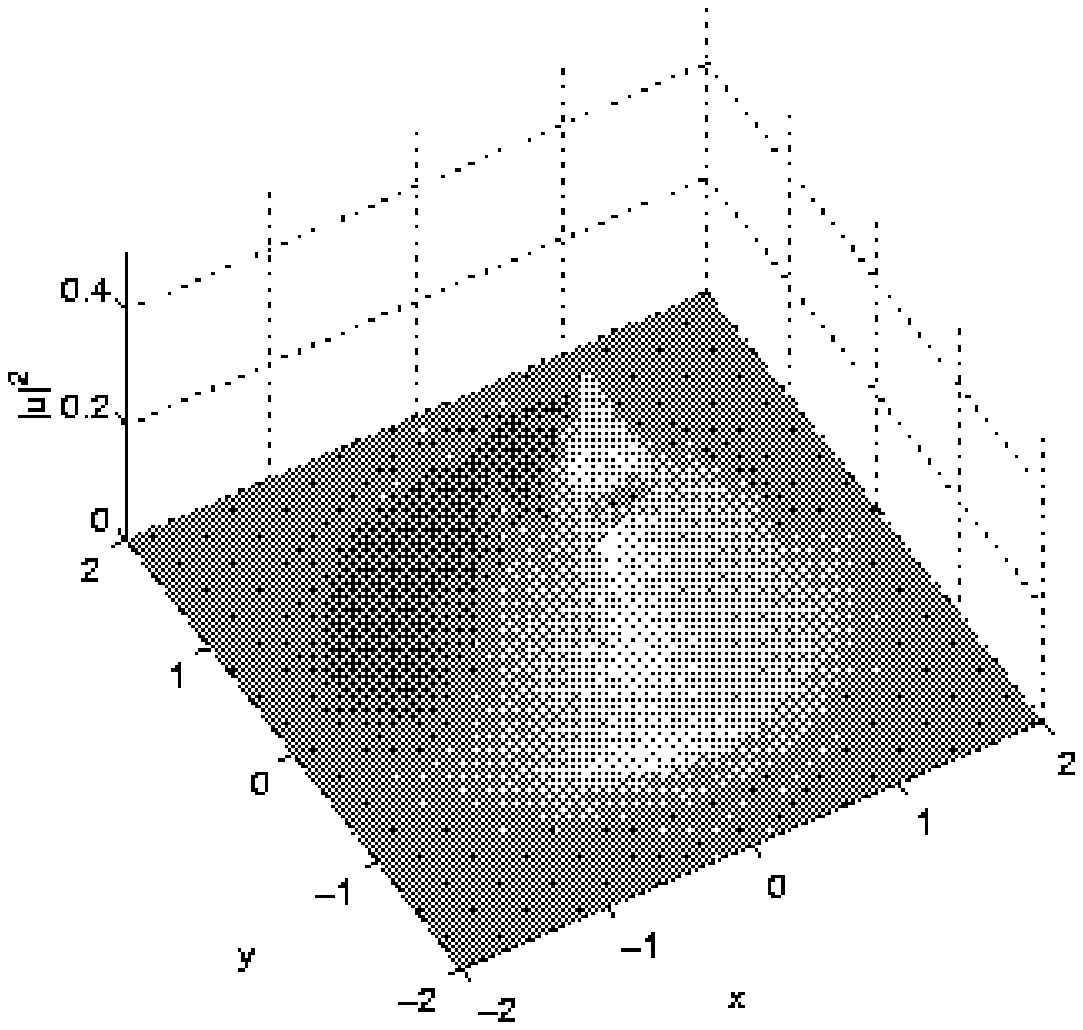,height=8cm,width=8cm,angle=0} 
b)\psfig{figure=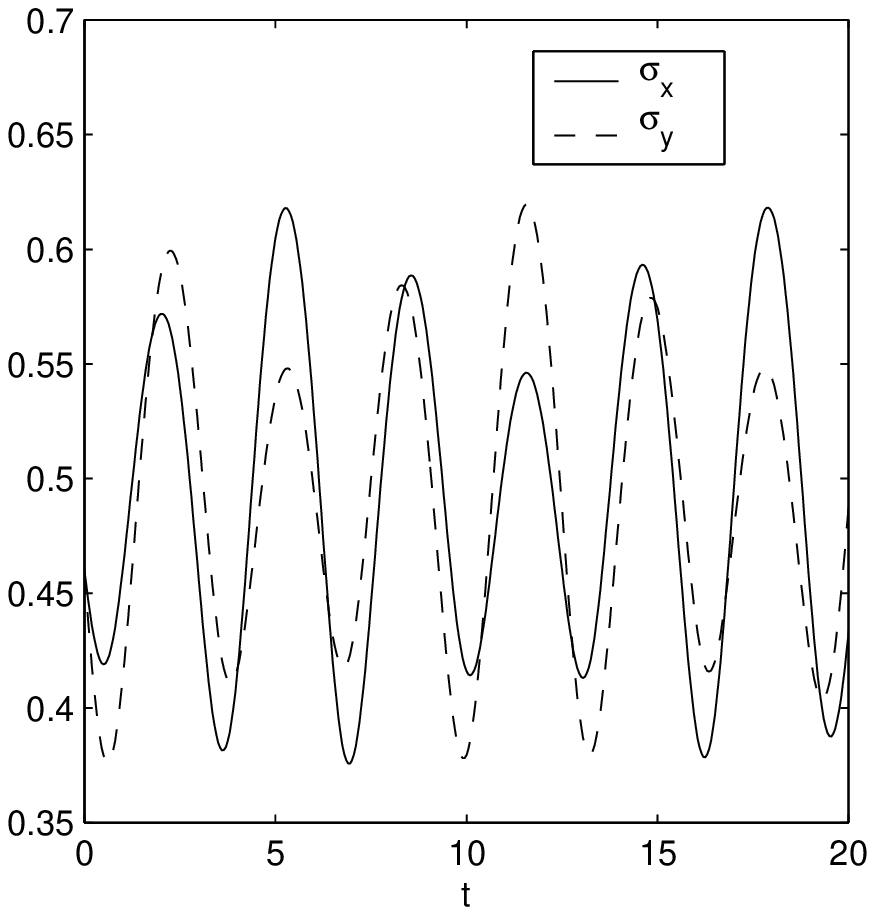,height=7cm,width=6cm,angle=0}} 
  Figure 5:  Numerical results in  Example 3 for case III. \quad 
a). Surface plot of the position density at $t=40$; 
b). widths of the condensate as a function of time. 
 
\vspace{2cm} 
\centerline{a)\psfig{figure=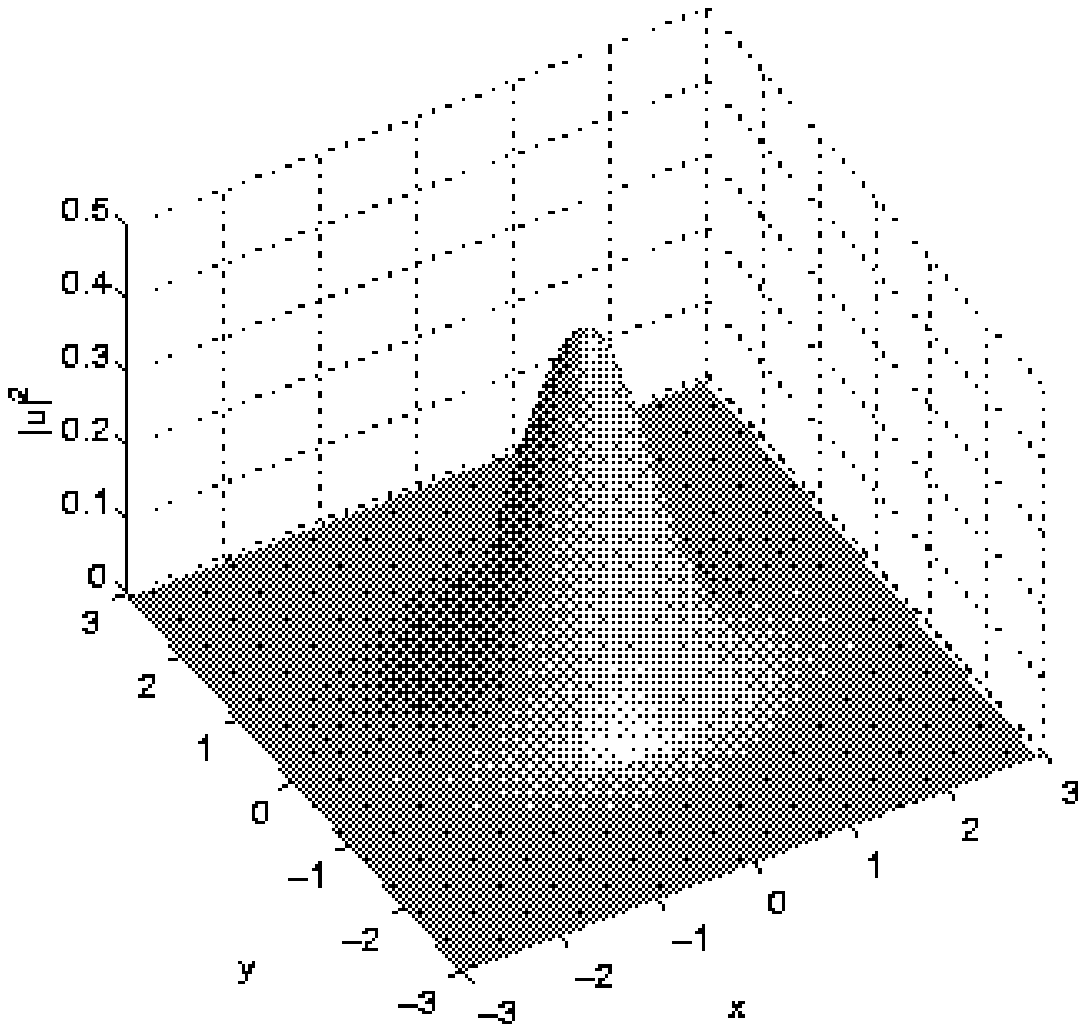,height=8cm,width=8cm,angle=0} 
b)\psfig{figure=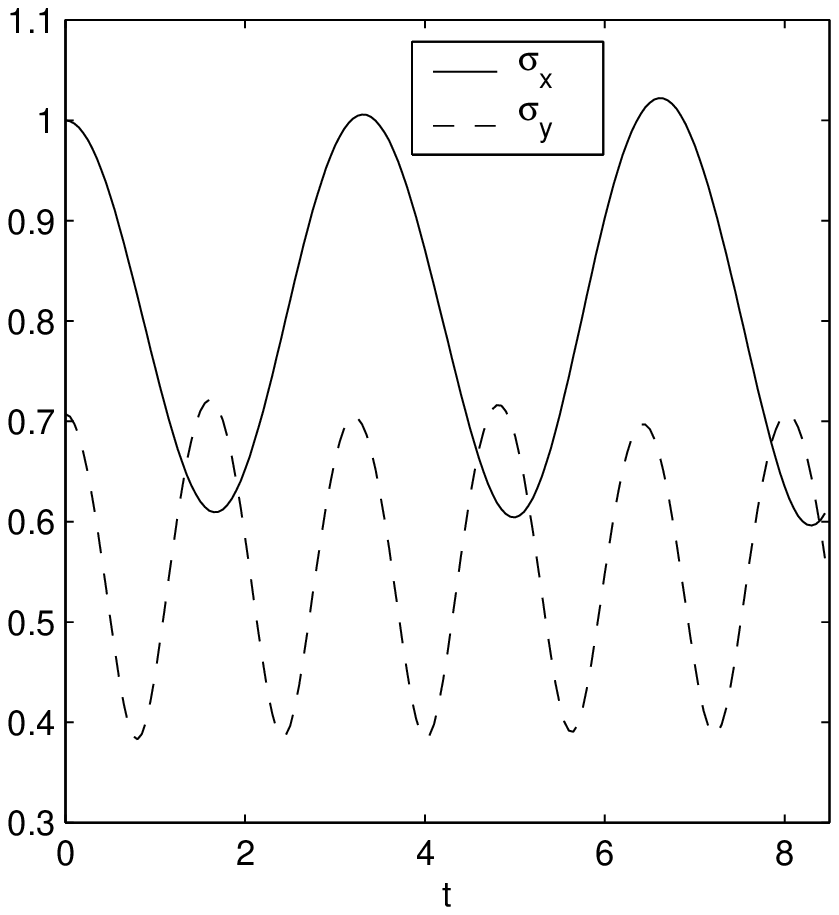,height=7cm,width=6cm,angle=0}} 
 
  Figure 6:  Numerical results in  Example 3 for case IV. \quad 
a). Surface plot of the position density at $t=6.8$; 
b). widths of the condensate as a function of time. 
 
\end{figure}

\newpage 
\begin{figure}[htb] 
\centerline{a).\psfig{figure=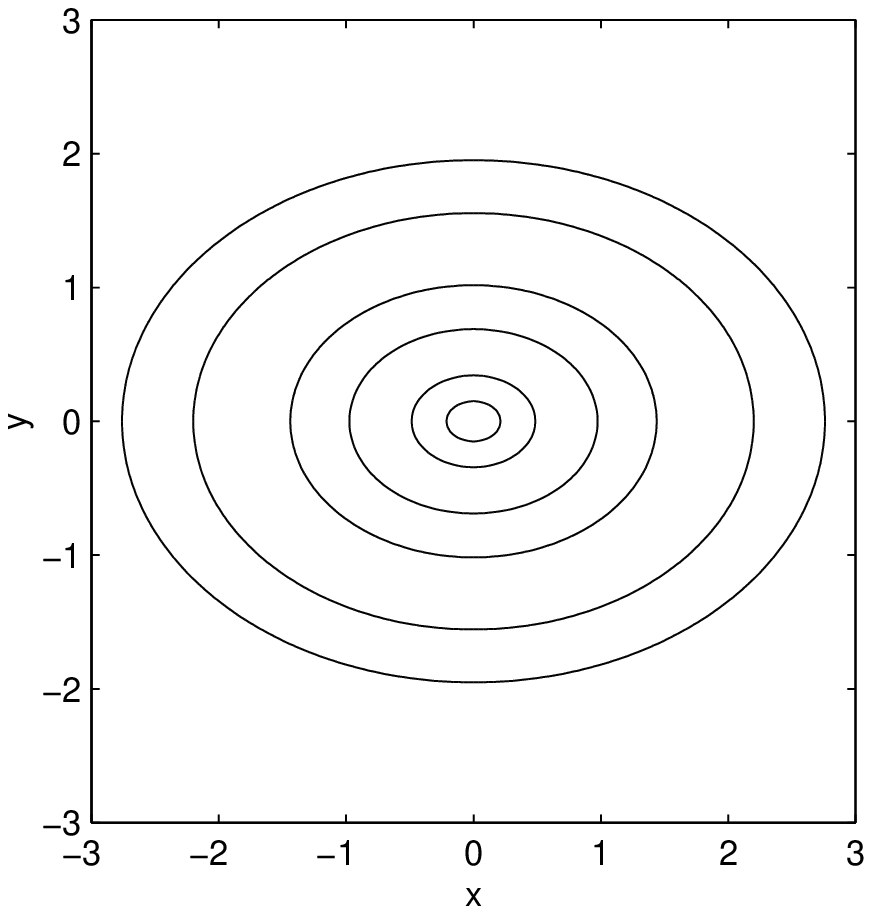,height=6.5cm,width=6.5cm,angle=0} 
\quad  b).\psfig{figure=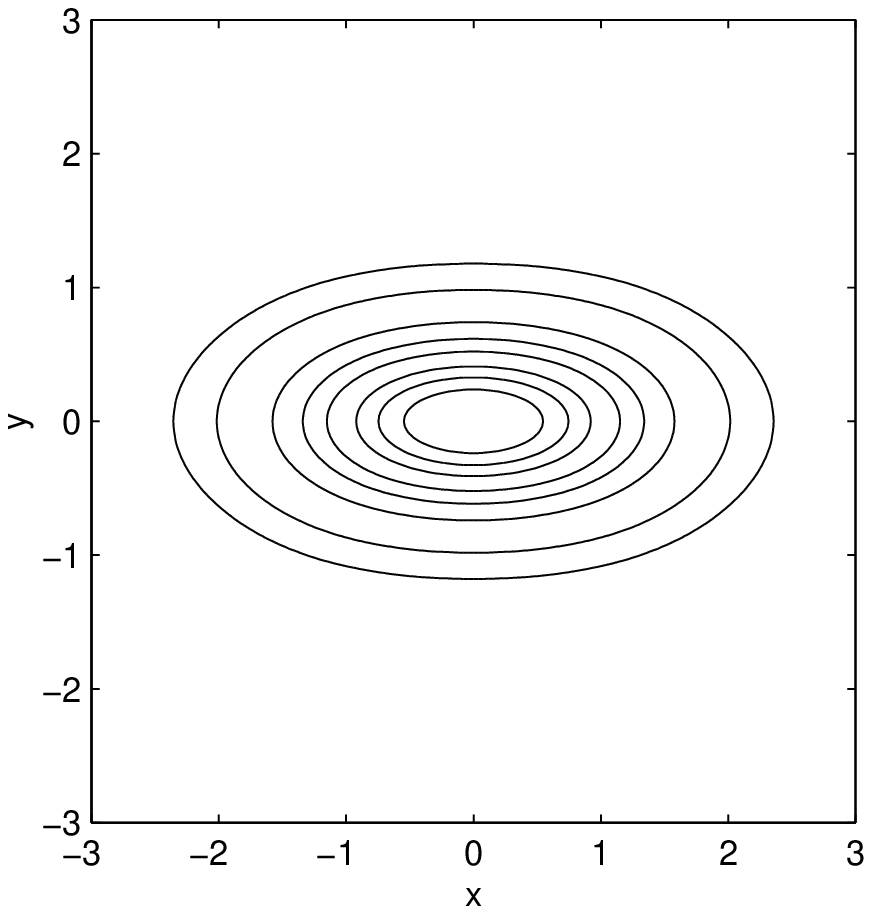,height=6.5cm,width=6.5cm,angle=0} } 
 \centerline{c).\psfig{figure=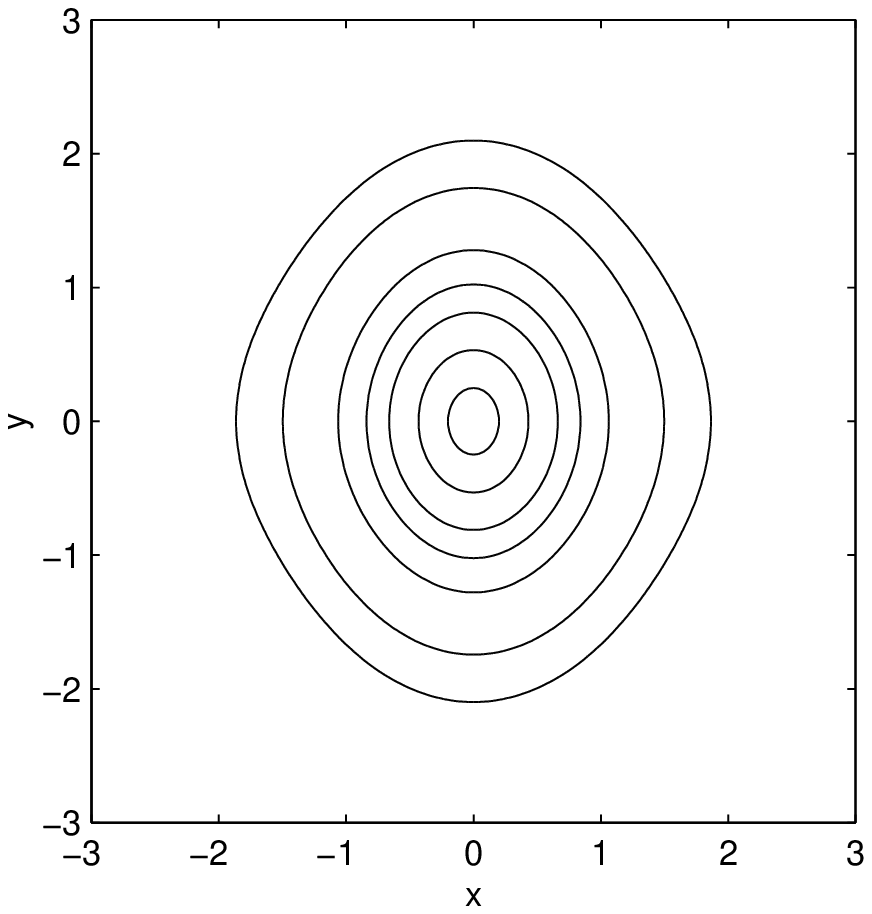,height=6.5cm,width=6.5cm,angle=0} 
\quad  d). \psfig{figure=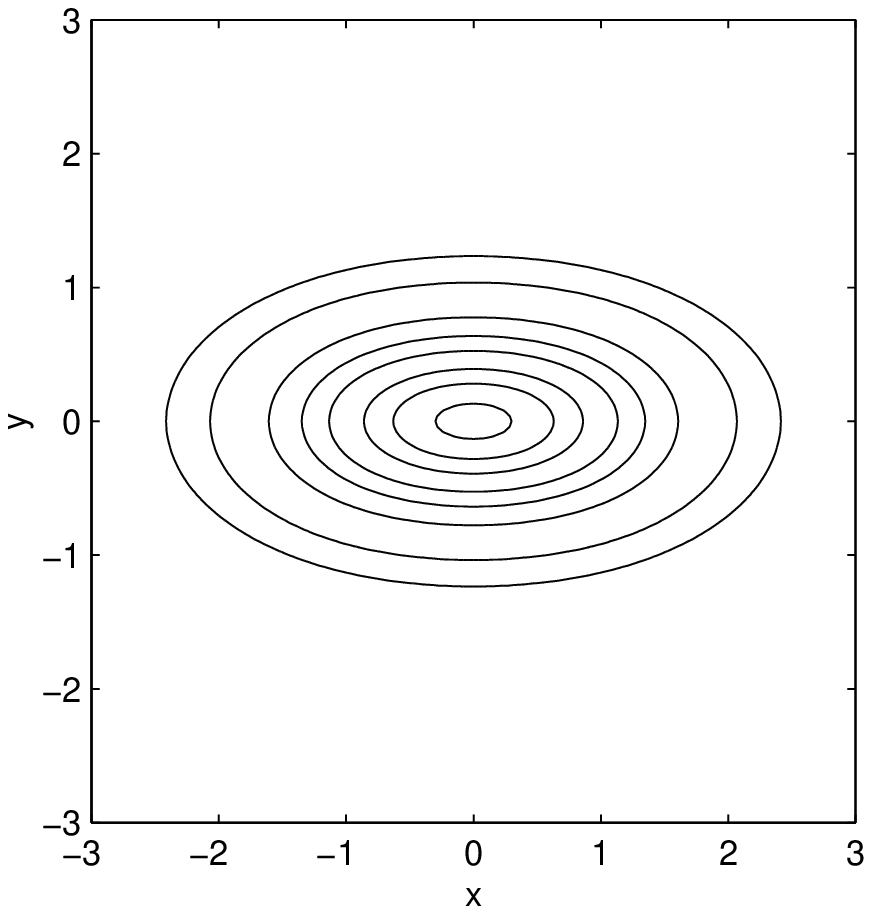,height=6.5cm,width=6.5cm,angle=0} } 
\centerline{e).\psfig{figure=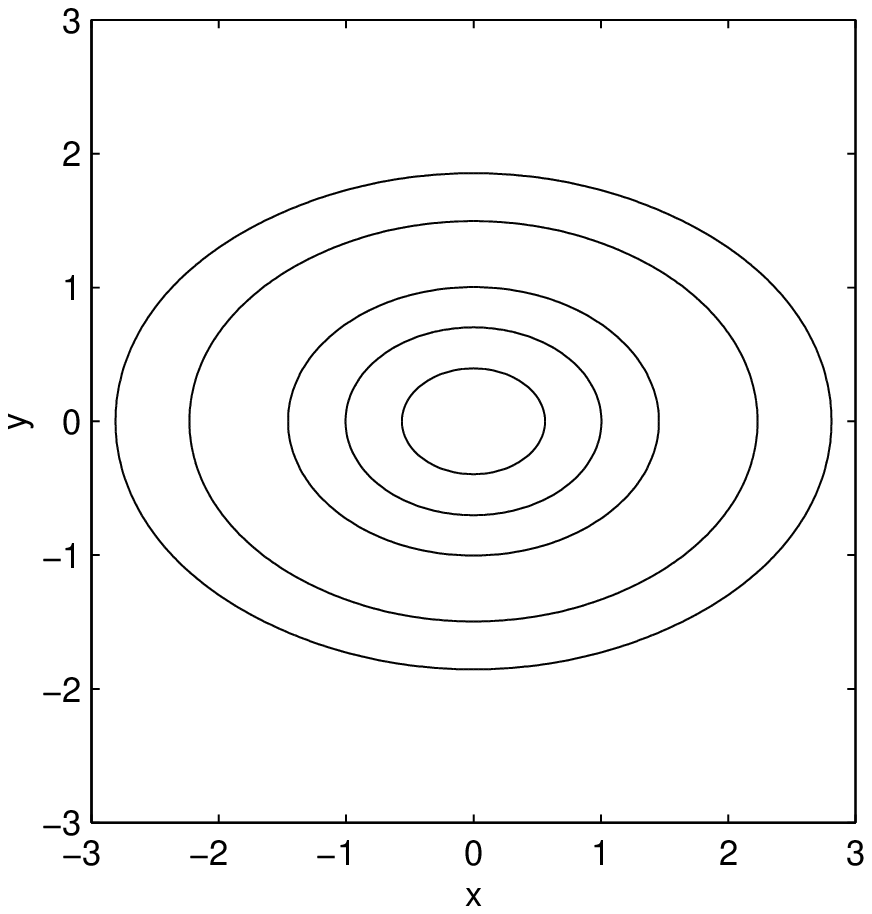,height=6.5cm,width=6.5cm,angle=0} 
\quad   f). \psfig{figure=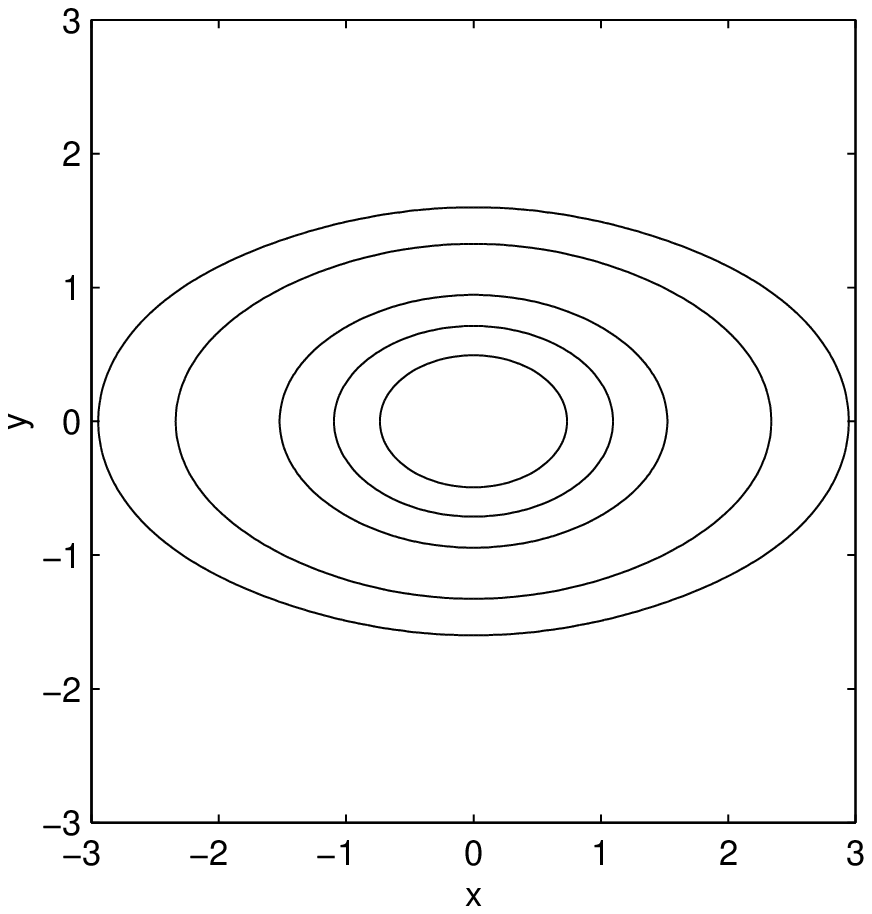,height=6.5cm,width=6.5cm,angle=0} } 
 
 Figure 7:  Contour plots of the position density at different times 
in  Example 3 for case IV. 
a). $t=0.0$, b). $t=0.85$, c). $t=1.7$, 
d). $t=2.55$, e). $t=3.4$, f). $t=6.8$. 
\end{figure} 
 
\newpage 
 \begin{figure}[htb] 
\centerline{a)\psfig{figure=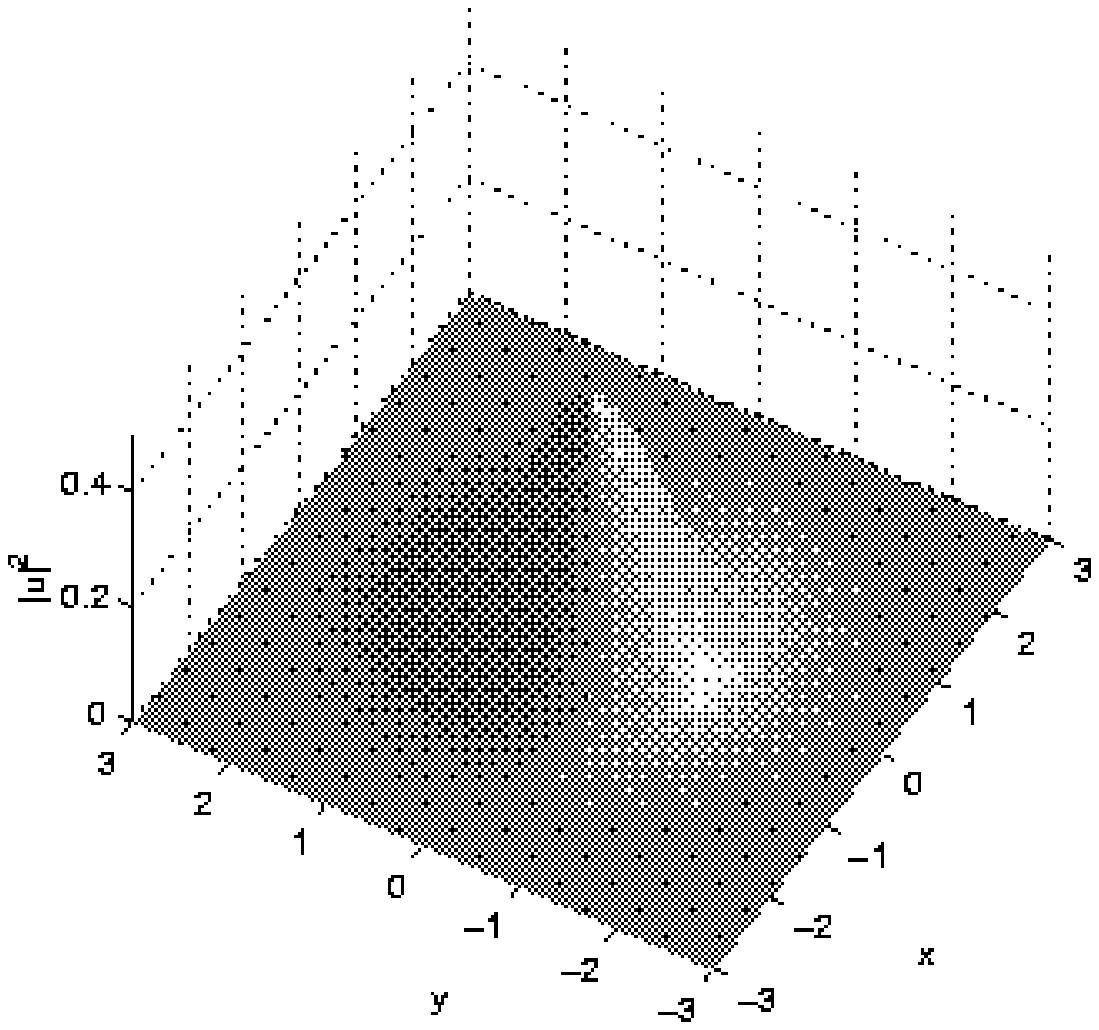,height=8cm,width=8cm,angle=0} 
b)\psfig{figure=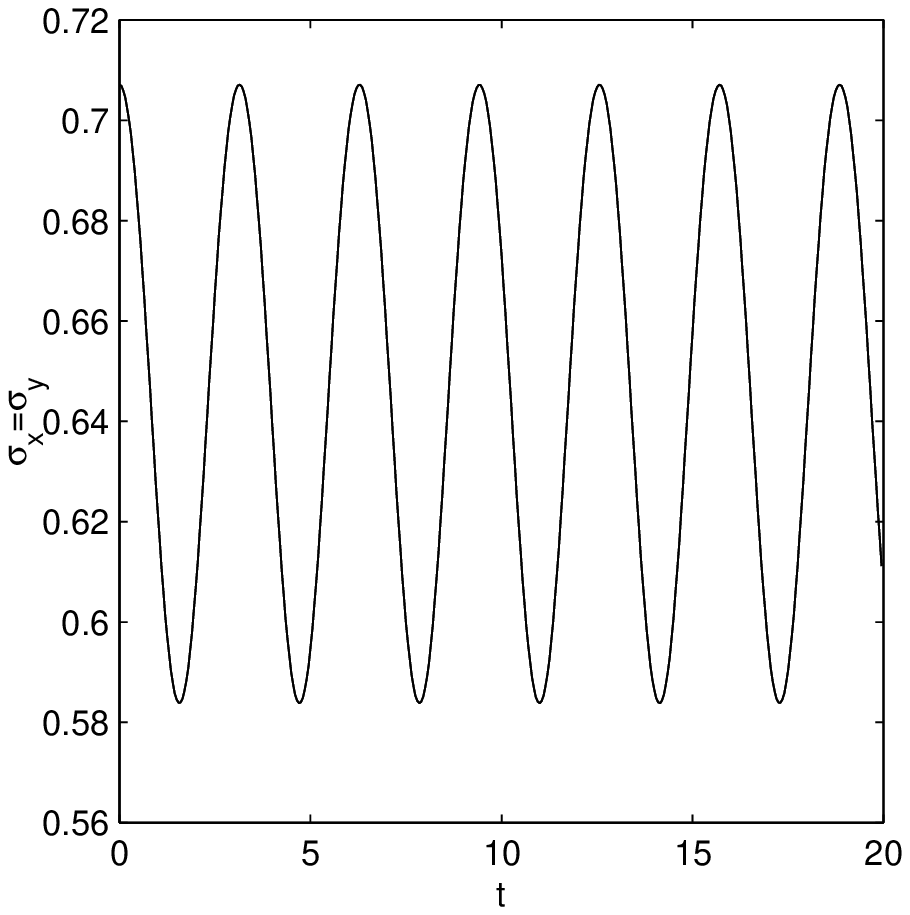,height=7cm,width=6cm,angle=0}} 
 
  Figure 8:  Numerical results in  Example 4 for case I. \quad 
a). Surface plot of the position density at $t=40$; 
b). widths of the condensate as a function of time. 
 
\vspace{2cm} 
 
\centerline{a)\psfig{figure=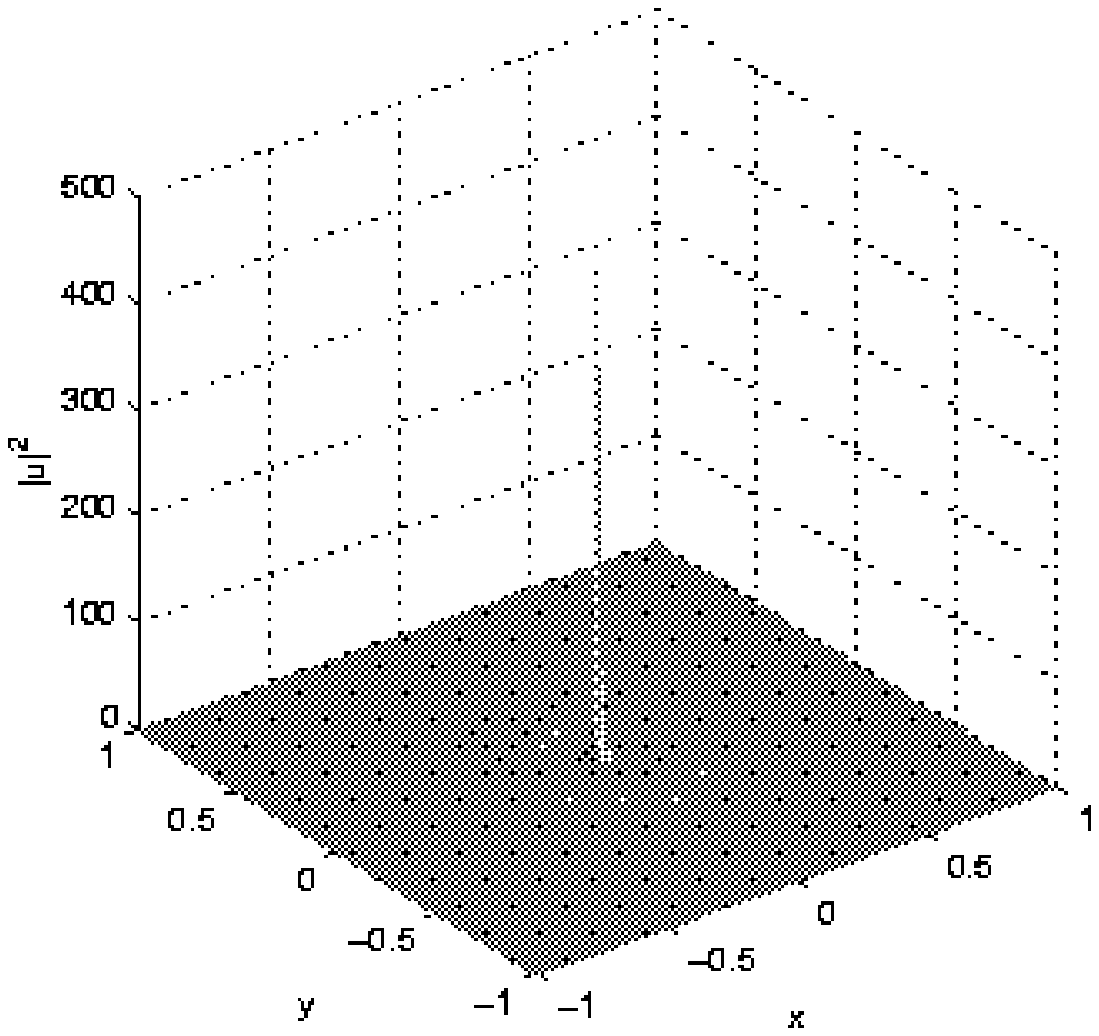,height=8cm,width=8cm,angle=0} 
b)\psfig{figure=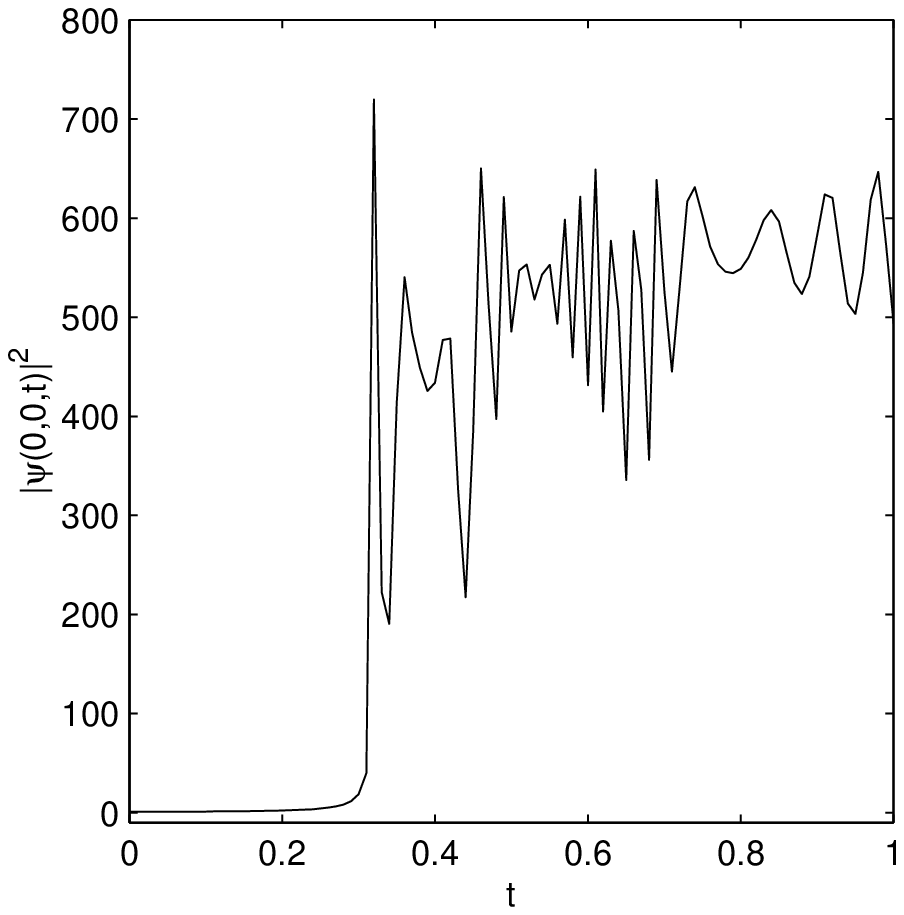,height=7cm,width=6cm,angle=0}} 
 
  Figure 9: Numerical results in  Example 4 for case II. \quad 
a). Surface plot of the position density at $t=0.5$; 
b). peak of the position density $|\psi(0,0,t)|^2$ 
 as a function of time. 
 
\end{figure} 
 
 \clearpage 
\newpage

\newpage 
 \begin{figure}[htb] 
\centerline{a)\psfig{figure=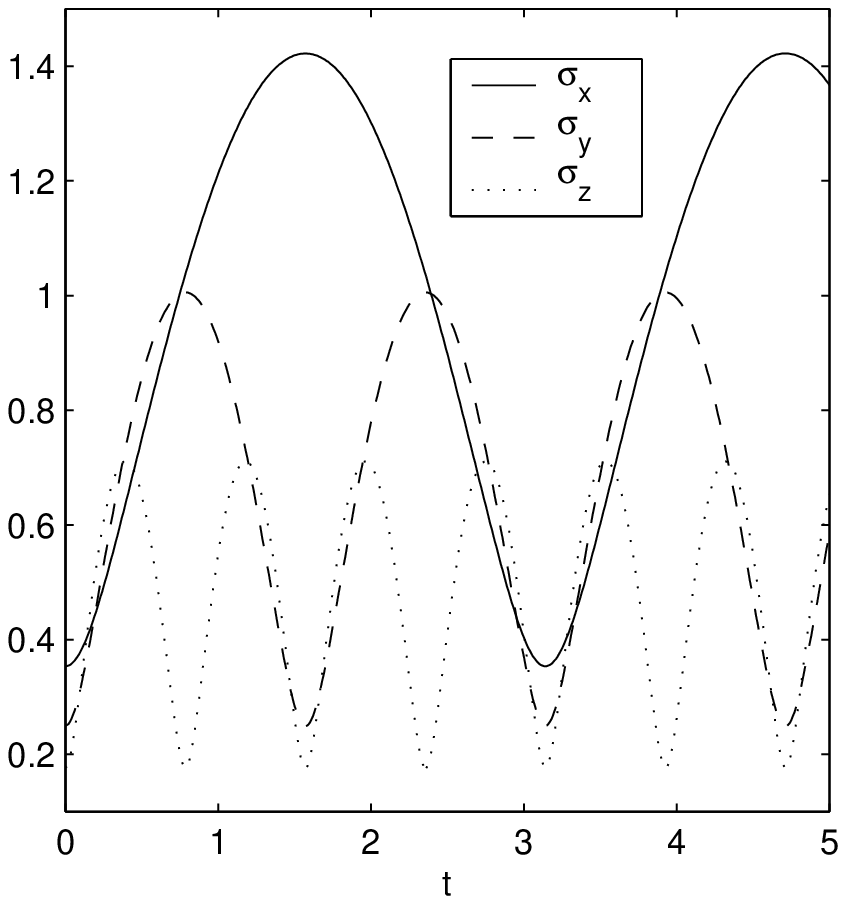,height=7cm,width=7cm,angle=0} 
b)\psfig{figure=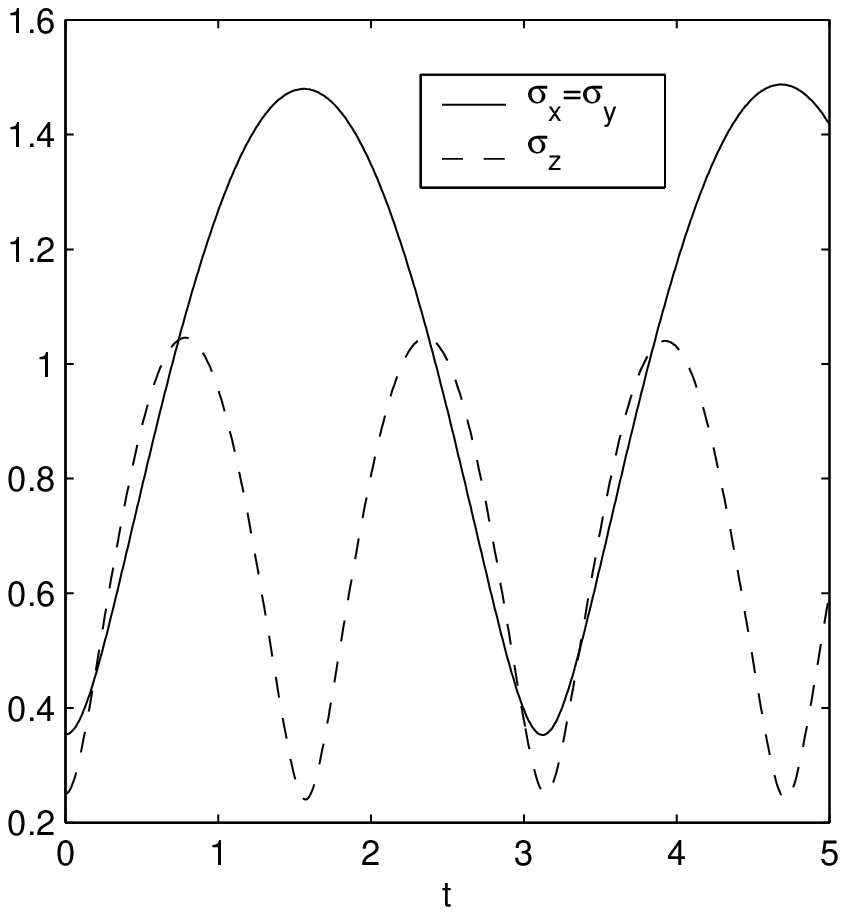,height=7cm,width=7cm,angle=0}} 
 
  Figure 10:  Widths of the condensate as a function of time in 
 Example 5. \quad a). For case I; b). for case II. 
 
\vspace{2cm} 
 
\centerline{a)\psfig{figure=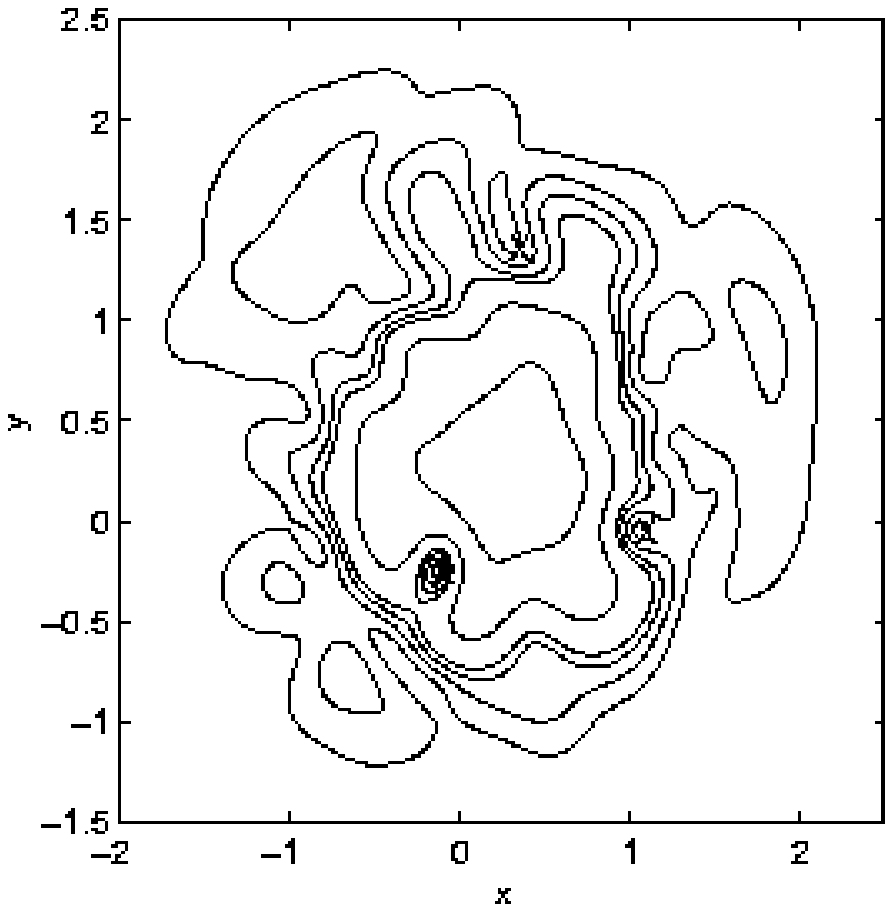,height=7cm,width=7cm,angle=0} 
b)\psfig{figure=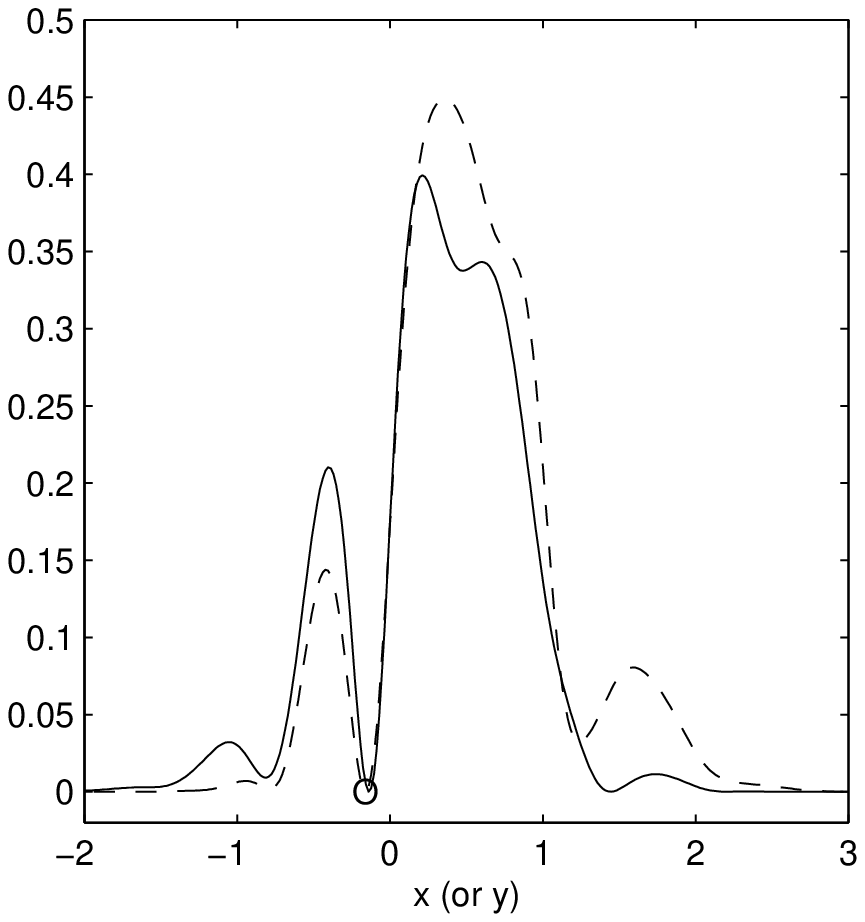,height=7cm,width=7cm,angle=0} 
} 
 
  Figure 11:  Numerical results at time $t=12\pi$ in Example 6. 
 \quad a). Contour plot of the density function $|\psi|^2$ 
(three vortices were identified and labelled by 'X'); 
b). $x,y$-sectional plot at a vortex (center of it is labelled 
by 'O') located at $(-0.141, -0.229)$; 
\  '---': $|\psi(x,-0.229,12\pi)|^2$, '- - - ': 
$|\psi(-0.141,y+0.088,12\pi)|^2$. 
\end{figure}

\end{document}

\newpage

\begin{figure}[htb] 
 
\end{figure}

\clearpage 

\clearpage 
\clearpage 

